\pdfoutput=1
\documentclass[manuscript]{acmart}
\usepackage[normalem]{ulem}
\useunder{\uline}{\ul}{}
\usepackage{booktabs}
\usepackage{tabularx} 
\usepackage{multirow}
\usepackage{color}
\usepackage{colortbl} 
\usepackage{fp} 
\usepackage{pgf} 
\usepackage{subcaption}

\AtBeginDocument{%
  }

\setcopyright{acmlicensed}
\copyrightyear{2018}
\acmYear{2018}
\acmDOI{XXXXXXX.XXXXXXX}

\acmConference[Conference acronym 'XX]{Make sure to enter the correct
  conference title from your rights confirmation emai}{June 03--05,
  2018}{Woodstock, NY}
\acmISBN{978-1-4503-XXXX-X/18/06}

\begin{document}

\title{In Shift and In Variance: Assessing the Robustness of HAR Deep Learning Models against Variability}

\author{Azhar Ali Khaked}
\affiliation{%
  \institution{Department of Electrical and Computer Engineering, Concordia University}
  \city{Montreal}
  \state{QC}
  \country{Canada}
}

\author{Nobuyuki Oishi}
\author{Daniel Roggen}
\affiliation{%
  \institution{University of Sussex}
  \city{Brighton}
  \country{UK}}

  \author{Paula Lago}
\affiliation{%
  \institution{Department of Electrical and Computer Engineering, Concordia University}
  \city{Montreal}
  \state{QC}
  \country{Canada}
}







\renewcommand{\shortauthors}{Ali Khaked et al.}

\begin{abstract}

Human Activity Recognition (HAR) using wearable inertial measurement unit (IMU) sensors can revolutionize healthcare by enabling continual health monitoring, disease prediction, and routine recognition. Despite the high accuracy of Deep Learning (DL) HAR models, their robustness to real-world variabilities remains untested, as they have primarily been trained and tested on limited lab-confined data. In this study, we isolate subject, device, position, and orientation variability to determine their effect on DL HAR models and assess the robustness of these models in real-world conditions.

We evaluated the DL HAR models using the HARVAR and REALDISP datasets, providing a comprehensive discussion on the impact of variability on data distribution shifts and changes in model performance. Our experiments measured shifts in data distribution using Maximum Mean Discrepancy (MMD) and observed DL model performance drops due to variability. We concur that studied variabilities affect DL HAR models differently, and there is an inverse relationship between data distribution shifts and model performance. The compounding effect of variability was analyzed, and the implications of variabilities in real-world scenarios were highlighted. MMD proved an effective metric for calculating data distribution shifts and explained the drop in performance due to variabilities in HARVAR and REALDISP datasets.

Combining our understanding of variability with evaluating its effects will facilitate the development of more robust DL HAR models and optimal training techniques. Allowing Future models to not only be assessed based on their maximum F1 score but also on their ability to generalize effectively.

\end{abstract}



\received{20 February 2007}
\received[revised]{12 March 2009}
\received[accepted]{5 June 2009}

\maketitle

\section{Introduction}
Most human activities involve predictable physical motions. Inertial Measurement Unit (IMU) sensors in smart devices, such as smartphones and watches, detect these movements and can be utilized by Human Activity Recognition (HAR) models for activity classification. Conventional Machine Learning (ML) models have been applied to HAR but require manual feature engineering and domain expertise. In contrast, Deep Learning (DL) models such as Convolution Neural networks (CNNs) have been shown to automate feature extraction~\cite{DLDoesntNeedFeatureEngineering} from input data in images~\cite{CNNFeature} and sound~\cite{CNNSound}. This automation has driven the development of various DL HAR models, showing promising results in activity classification from wearable IMU sensors. DL HAR models like \citep{dLHARoutperform1} and \citep{dLHARoutperform2} offer excellent performance on existing HAR datasets. However, these models have been trained and tested on datasets typically collected in laboratory settings. In these datasets, participants perform activities restrictively, eliminating any variability that might arise from human or hardware changes. While DL HAR models demonstrate excellent performance on constrained and small datasets, their robustness to various forms of variability remains untested.

Variabilities occur due to device changes, wearing habits, and varying users. Any physical change to an IMU sensor that alters its measurements can be considered variability~\cite{introdevice}. Device variability results from hardware variations, including sensor sensitivity, sampling rate, range, and noise~\cite{devicevarintro}. Wearing variability arises from changes in the position or orientation of the sensor worn by the user~\cite{wearingVarIntro}. Subject variability arises due to users performing actions in varying ways~\cite{subjectvar1, subvarmotion1}. 

Variability induces a distribution shift in the data used to train and test DL HAR models. A DL HAR model must be robust to these variabilities to be reliable in healthcare, lifestyle, and pervasive health monitoring applications. While it is known that data distribution affects the performance of DL models in general, the effects of the specific variabilities affecting IMU sensors in the performance of DL HAR models are unknown. Assessing DL HAR models on data with no variability provides a limited understanding of their performance and robustness. Therefore, it is crucial to understand how distribution shifts in data due to variability affect the performance of DL HAR models and how to incorporate variability into their performance evaluation.

This empirical study makes the following contributions to our understanding of the effects of variability in DL HAR models:
\begin{itemize}
    \item We create a dataset designed specifically to study the effect of variability in HAR, with two device types in six positions and varying orientations at a singular position~(Section~\ref{sec:harvar}). 
    \item We quantify the effect of human, wearing, and device variabilities caused by subject, orientation, position, and hardware changes~(Section~\ref{sec:model_performance}).
    \item We measure the relationship between the impact of variability and the shift in data distribution by using Maximum Mean Discrepancy (MMD)\cite{gretton2012kernel}~(Section~\ref{sec:mmd_var}).
\end{itemize}

Following this approach, we aim to understand the data shift due to device and wearing variability and study their effects on DL HAR models.


\section{Related Work}

Real-life variation arises from human or hardware changes, leading to distribution shifts. The impact of distribution shifts on DL models has been studied in domains such as image recognition and audio processing and has shown adversarial effects.

In the audio domain, the input signal is continuous and similar to IMU sensor data. \citet{soundDistributionShift} studied the effect of distribution shift on DL models used to classify industrial sound and found performance drops to be related to the changes in the distribution of the data. Distribution shifts accounted for 9-10\% drops in the DL model performances.

\citet{imageDistributionShift} assess the robustness of image classification models by evaluating models not based on their accuracy but by comparing the change in performance when testing the model with two test sets: One test had no distribution shift, and another test set had a distribution shift. Even with extensive training data, they found that DL models show a high susceptibility and corrupted classifications due to distribution shifts. 

While \citep{imageDistributionShift} measures the performance drop by changing the test set, the study by~\citet{wilds} recommends measuring the performance drop by testing on the same distribution to isolate the effect of the distribution shift. Therefore, this research will use the latter approach to evaluate the effects of variability.

Subject variability is a common challenge in motion-based HAR using wearable sensors, as individuals perform activities in different ways \cite{subjectAndMotionVarWithVariantAndInvariantSVM}. \citet{subVar} investigated the impact of subject variability on traditional ML HAR classification models and Convolutional Neural Networks (CNNs). They trained their models on data collected from adults aged 19-48 and tested them on data from elderly individuals. The study found that ML models experienced an average performance drop of 9.75\%, while CNNs showed a larger decline of 12\%, indicating that DL models may be more vulnerable to variability. Additionally, they observed changes in the features extracted for classification, suggesting that the age-related shift in data distribution affected model performance. Our study does not study subject variability between groups but for each individual against the rest using leave one subject out (LOSO) Cross-validation.

Orientation Variability in HAR has been studied in~\citet{orientation1}, where significant but varying drops in the performance of ML models were noted when the test data was subjected to random rotation. They found that some datasets experienced a 30\% drop in accuracy due to rotation, while others showed no change. The paper did not explain this varying effect of orientation variability but focused on methods to reduce its impact.

\citet{orientationVarReductionUsingReference} studied the effect of orientation change on CNNs. The baseline performance is obtained by training and testing the network with the original data from six public HAR datasets. 45$^\circ$ of orientation changes were induced via matrix transformation on the test set to measure the performance after the orientation change. They found that rotation transformation caused the model accuracy to drop by 2\% to 11\%, depending on which dataset was being used to train the ML models.

\citet{myOldPaper} did not artificially induce orientation changes using transformations but instead used data from two sensors placed at the same location on participants, angled 45$^\circ$ relative to each other, capturing realistic shifts in data distribution due to orientation changes. Following \citet{wilds}, they evaluated the impact of orientation variability using the same test set, training the model both with and without this variability. The performance difference between the two models averaged 2\%, with variation across participants. They observed a negative correlation between performance drop and changes in MMD values, though their study was limited to 8 participants, with 2 not exhibiting this negative relationship.

Wearing variability for devices such as earbuds, which are fixed in position, can only induce orientation variability, as shown by~\citet{wearingvar1}. They found that orientation changes increased the Euclidean distance between the IMU data collected from different sessions for the same person and between different people and their habits. Positional variability due to sensor placement was studied on animals in~\citet{animalvar}, where changing the sensor position from the back to the neck on dogs and horses led to a significant drop in the performance of unsupervised models for Animal Activity Recognition (AAR).

\citet{deviceVarHHAR} investigated the effect of device variability on HAR classification by training a Recurrent Neural Network model using smartphones and testing it with data from smartwatches. This resulted in an accuracy of 45\%, nearly half the accuracy of other scenarios based on participant-wise train-test splits. This highlighted the impact of device variability, although it was mixed with the effect of position variability, as smartwatches were worn on the wrist while smartphones were worn on the waist.

Orientation, position, and device variability have been identified as a problem in HAR \cite{positionVar1, devicevarintro}, but they have either not been studied in isolation or not in the context of DL models. This has limited understanding of how these variabilities affect data distribution shifts and model performance. We address this gap by analyzing the performance changes of DL-HAR models under isolated and combined variabilities and explaining the performance drop using MMD to measure data distribution shifts.


\section{Method}
As mentioned before, variability can be subdivided into subject, device, and wearing variability. 
Every variability induces a shift in the distribution of the data, leading to a change in the performance of DL HAR models. In this section, we explain the deep learning models used, the dataset collected for the evaluation, and the experimental protocol used to evaluate the effects of variability in DL-HAR models. We then explain the training settings for models and the performance evaluation selected. 

\subsection{Deep Learning Models for HAR evaluated}

Many DL HAR models have been proposed recently, and most can be categorized as homogeneous or hybrid. Homogeneous models like those in \cite{simres4} and \cite{simres5} exclusively use deep CNN or RNN architectures. 
In contrast, hybrid models combine CNNs with sequential neural networks, such as RNNs \cite{simres3}, LSTMs \cite{deepconvlstm}, and GRUs \cite{simres1}. Research, such as \citet{deepconvlstm}, has demonstrated that hybrid models generally exhibit superior all-around performance compared to homogeneous models. For instance, \citet{deepconvlstm} tested HAR classification using their proposed Deep CNN with LSTM approach and found an average improvement of 6\% over architectures that exclusively use deep CNNs. This improvement can be attributed to the complementary behaviors of CNNs and sequential networks. CNNs excel at extracting local features from the input and capturing spatial patterns. In contrast, sequential networks such as RNNs, LSTMs, and GRUs extract temporal features over the entire input window, identifying patterns and dependencies across time. Combining these strengths, hybrid models effectively leverage local and temporal information.

For this study, we have chosen to evaluate the robustness of three hybrid DL-HAR models: DeepConvLSTM \cite{deepconvlstm}, TinyHAR \cite{tiny}, and Attend and Discriminate \cite{attanddisc}. These models were selected due to their unique feature extraction and temporal information processing approaches, as detailed in Table \ref{tab:har_dl_table}. DeepConvLSTM \cite{deepconvlstm} serves as a representative model for most hybrid DL HAR models, combining CNN with LSTM to extract local and temporal features. Originally, DeepConvLSTM used two LSTM layers, as proposed by \citet{WhyDCLhasTwoLSTMLayers}, but \citet{newdeepconvlst} demonstrated that a single LSTM layer performs better in most cases, which is why we will use a shallow DeepConvLSTM in this study. Both Attend and Discriminate \cite{attanddisc} and TinyHAR \cite{tiny} employ attention mechanisms to improve feature extraction, with TinyHAR additionally optimizing the model to be lightweight.

\begin{table}[ht]
\begin{tabular}{p{4cm}p{5cm}p{5cm}p{5cm}}
\toprule 
\textbf{DeepConvLSTM} & \textbf{Attend and Discriminate} & \textbf{TinyHAR}
\\
\midrule
\vspace{1pt}
{\ul {Feature Extraction: (Extraction of local or short time features)}}\\
\vspace{1pt}
Four 1-dimensional convolution layers with a kernel size of 5, a stride of 2, and 64 filters are used to extract local features from input data. & Local feature extraction is done in two steps: First, the data is processed through 4 one dimensional convolution layers with a kernel size of 5, stride 2, and 64 filters. Second, a transformer encoder block comprising a self-attention and two fully connected feed-forward layers encode channel interaction. Where channels are the readings from various sensor modalities. & Local feature extraction is done in three steps: First, the data is processed through 4 one dimensional convolution layers with a kernel size of 5, stride 2, and 20 filters. Second, a transformer encoder block comprising a self-attention and two fully connected feed-forward layers encode channel interaction. Third, a fully connected layer fuses the cross-channel interaction information. 
\\
\vspace{6pt}
{\ul {Temporal Information Extraction: (Extraction of features over the entire time window)}}\\
\vspace{2pt}
A single LSTM layer with 128 cells is used to extract temporal features.& Temporal information is extracted in two steps: First, a single GRU layer with 128 cells extracts temporal features. Second, a self-attention layer is used to highlight important temporal features.& Temporal information is extracted in two steps: First, a single LSTM layer with 40 cells extracts temporal features. Second, a self-attention layer is used to highlight important temporal features.\\
\bottomrule
\end{tabular}
\caption{The three SOTA DL HAR models used for this study and their key architectural differences.}
\label{tab:har_dl_table}

\end{table}


\subsection{The Human Activity Recognition Variability (HARVAR) Dataset} \label{sec:harvar}
We collected a dataset highlighting the effect of variability in HAR by using multiple sensors simultaneously in varying positions and orientations and with two types of devices: Empatica Embrace Plus and Bluesense~\cite{bluesense}. 

\begin{figure}
    \centering
    \includegraphics[width=0.75\linewidth]{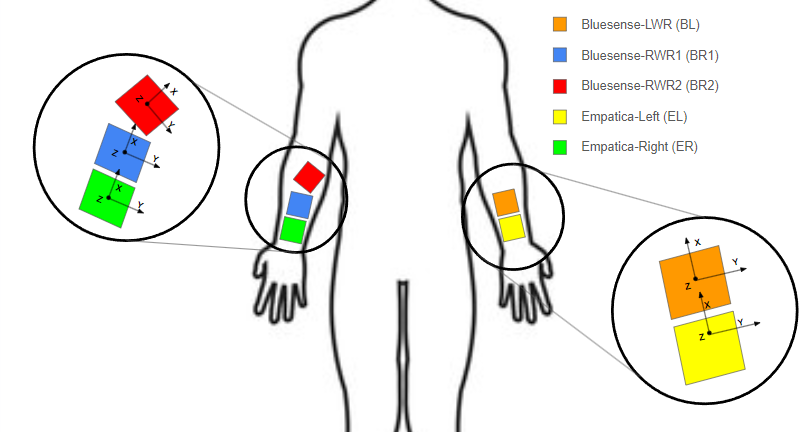}
    \caption{Placement of sensors in HARVAR data collection. The Empatica Embrace Plus and Bluesense sensors are placed in the same coordinate system, and their axis is marked. BR2, marked as red, is tilted across the Z-axis at 45-degree of rotation. In this diagram, the person is facing towards the reader}
    \label{fig:harvar_placement}
\end{figure}

Figure~\ref{fig:harvar_placement} shows how the devices were attached to each participant. On the right wrist, participants had 3 devices: 1 Empatica (ER),1 Bluesense (BR1) in the upright orientation, and 1 Bluesense (BR2) with a 45-degree rotation. In the left wrist, participants had 2 devices: 1 Empatica (EL) and 1 Bluesense (BL) in the upright orientation. This information is summarized in table \ref{tab:sensors_list}. The sensors ER and BR1 follow the same coordinate system; similarly, BL and EL follow the same coordinate system.

\begin{table}[ht]
\centering
\begin{tabular}{lllll}
\toprule
\textbf{Sensor Type} & \textbf{Sensor Position}  & \textbf{Sensor Name} & \textbf{Sensor Code} & \textbf{Sampling Frequency} \\
\midrule
Bluesense            & Right Wrist (No rotation) & bluesense-RWR1       & BR1 & 100Hz\\
Bluesense            & Right Wrist (45 rotation) & bluesense-RWR2       & BR2 & 100Hz\\
Bluesense            & Left Wrist                & bluesense-LWR        & BL & 100Hz\\
Empatica             & Right Wrist               & empatica-right       & ER & 64Hz\\
Empatica             & Left Wrist                & empatica-left        & EL & 64Hz\\
\bottomrule

\end{tabular}
\caption{The sensors used from the collection of the HARVAR dataset along with information on their sampling rate and placement.}
\label{tab:sensors_list}
\end{table}

Each participant performed two types of activities: treadmill walking and preparing a simple salad. 
The treadmill walks were conducted at speeds of 3.2 km/h, 4 km/h, 4.8 km/h, 5.6 km/h, and 6.4 km/h. 
Each participant walked for 2 minutes per speed, meaning every participant would walk for approximately 10 minutes. On the other hand, the average time spent in the salad preparation was 20 minutes per participant. By collecting data simultaneously from all sensors, we eliminate the variability introduced by human factors when the same activity is performed multiple times. During the treadmill walking phase of the experiment, the participants were not given any instructions on how to walk and were requested to walk in a way most comfortable and natural to them. Due to this, three participants chose to hold onto the support rails while walking on the treadmill, and others chose not to, as shown in Table~\ref{tab:participantsInfo}.

The dataset includes 16 participants from diverse age groups, with a mean age of 42 years and a standard deviation of 20 years. The data consists of 9 male participants, with a mean weight of 74 kg and a standard deviation of 13 kg, and 7 female participants, with a mean weight of 62 kg and a standard deviation of 13 kg. The collection took place under minimal restrictions to obtain natural IMU readings.
Since 30 minutes of labeled data were collected per participant, HARVAR has 8 hours of data.

\begin{table}[ht]
    \centering
    \begin{tabular}{ccccc}
    \toprule
        Age & Sex & Weight (Kgs) & ID & Holding Sidebar \\ \midrule
        59 & m & 83.9 & 1  & no \\ 
        74 & f & 65.7 & 2  & yes\\ 
        60 & f & 49.8 & 3  & no \\ 
        71 & m & 79.0 & 4  & yes\\ 
        61 & f & 55.3 & 5  & no \\ 
        71 & m & 64.8 & 6  & no \\ 
        26 & f & 73.0 & 7  & no \\ 
        25 & m & 72.5 & 8  & no \\ 
        26 & m & 61.0 & 9  & yes\\ 
        47 & m & 89.8 & 10 & no \\ 
        23 & f & 53.0 & 11 & no \\ 
        21 & m & 55.0 & 12 & no \\ 
        24 & f & 74.8 & 13 & no \\ 
        35 & f & 86.2 & 14 & no \\ 
        29 & m & 73.0 & 15 & no \\ 
        26 & m & 95.0 & 16 & no \\ 
        \bottomrule
    \end{tabular}
    \caption{Inforamtion about the 16 participants of HARVAR Dataset.}
    \label{tab:participantsInfo}
\end{table}


\subsection{Experimental Protocol for Model Robustness Evaluation}

\begin{figure}
    \centering
    \includegraphics[width=0.9\linewidth]{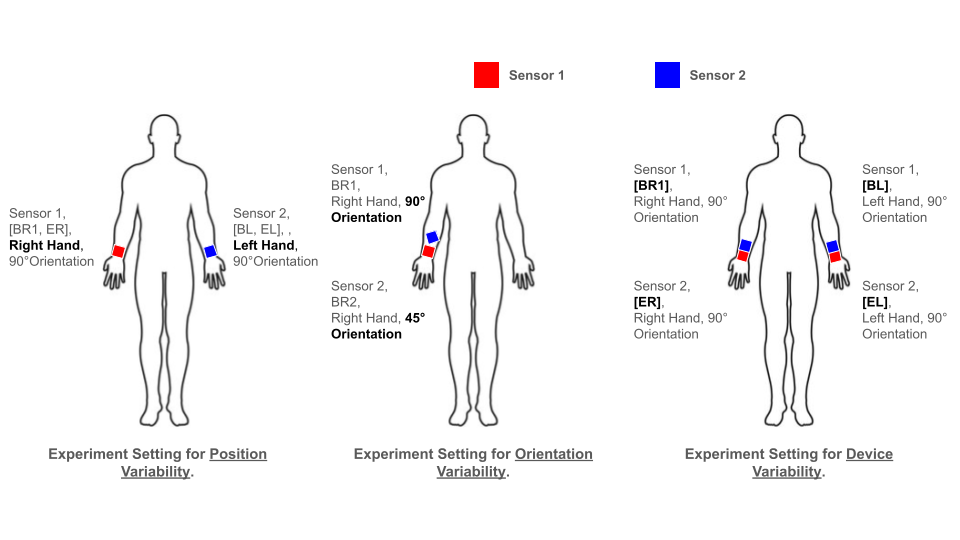}
    \caption{The experiment setting using the HARVAR dataset to evaluate the effect of device, position, and orientation variability. Where Sensor 1 and Sensor 2 are used in combination as a train-test pair to highlight variability. In these diagrams, the person is facing towards the reader.}
    \label{fig:harvar_exp_set}
\end{figure}

To understand the effect of each type of variability to be studied, we selected sensor pairs representing Device, Position, and Orientation Variability. We evaluated each model for selected pairs of sensors: once for the baseline scenario with no variability and once for the variability scenario. The pairs for position and orientation variability use the same type of device on different wrists (position) or the same wrist but with rotation (orientation), respectively. In contrast, we use the same position and orientation for device variability but different device types. These pairs are depicted in Table~\ref{tab:experiments} from rows 1 to 8 and Figure~\ref{fig:harvar_exp_set}. Notice that in each experiment, the test sensor is the same to isolate the effects of the variability following recommendations in~\cite{wilds}. Testing with a different sensor would combine the effects of the different testing distributions and the variability. We evaluate the effect of variability as the performance disparity, measured with F1-Score, between the two settings. 

The evaluation used cross-validation with a LOSO approach in each evaluation setting. For example, for experiment 1 in the variability setting, we trained each model using the data from the empatica-right sensor of Participants 2-16 and tested using the data from the empatica-left sensor of Participant 1 for the first fold. This also allows us to identify the subject variability between participants as in every baseline scenario as shown in Table~\ref{tab:experiments}, we can observe the difference in performance for each participant when their data is used for testing.

\begin{table}[ht]
    \centering
    \begin{tabular}{lllll}
        \toprule
         Exp. ID & Variability type & Train Sensor & Test Sensor& Setting \\
         \midrule
         \multirow{2}{*}{}1. & \multirow{2}{*}{}Position & empatica-right&empatica-left & variability\\
         && empatica-left & empatica-left & baseline\\ 
         \multirow{2}{*}{}2. & \multirow{2}{*}{}Position & empatica-left&empatica-right & variability\\
         && empatica-right & empatica-right & baseline\\ 
         \multirow{2}{*}{}3. & \multirow{2}{*}{}Position & BRW1 &BLW & variability\\
         && BLW & BLW & baseline\\ 
         \multirow{2}{*}{}4. & \multirow{2}{*}{}Position & BLW&BRW1& variability\\
         && BRW1& BRW1 & baseline\\ 
         \midrule
         \multirow{2}{*}{}5. & \multirow{2}{*}{}Device & BLW&empatica-left& variability\\
         && empatica-left& empatica-left & baseline\\ 
         \multirow{2}{*}{}6. & \multirow{2}{*}{}Device & empatica-left&BLW& variability\\
         && BLW& BLW & baseline\\
         \midrule
         \multirow{2}{*}{}7. & \multirow{2}{*}{}Orientation & BRW1&BRW2& variability\\
         && BRW2& BRW2 & baseline\\
         \multirow{2}{*}{}8. & \multirow{2}{*}{}Orientation & BRW2&BRW1& variability\\
         && BRW1& BRW1 & baseline\\
         \bottomrule
    \end{tabular}
    \caption{Experiments conducted}
    \label{tab:experiments}
\end{table}

\begin{figure}
    \centering
    \includegraphics[width=0.8\linewidth]{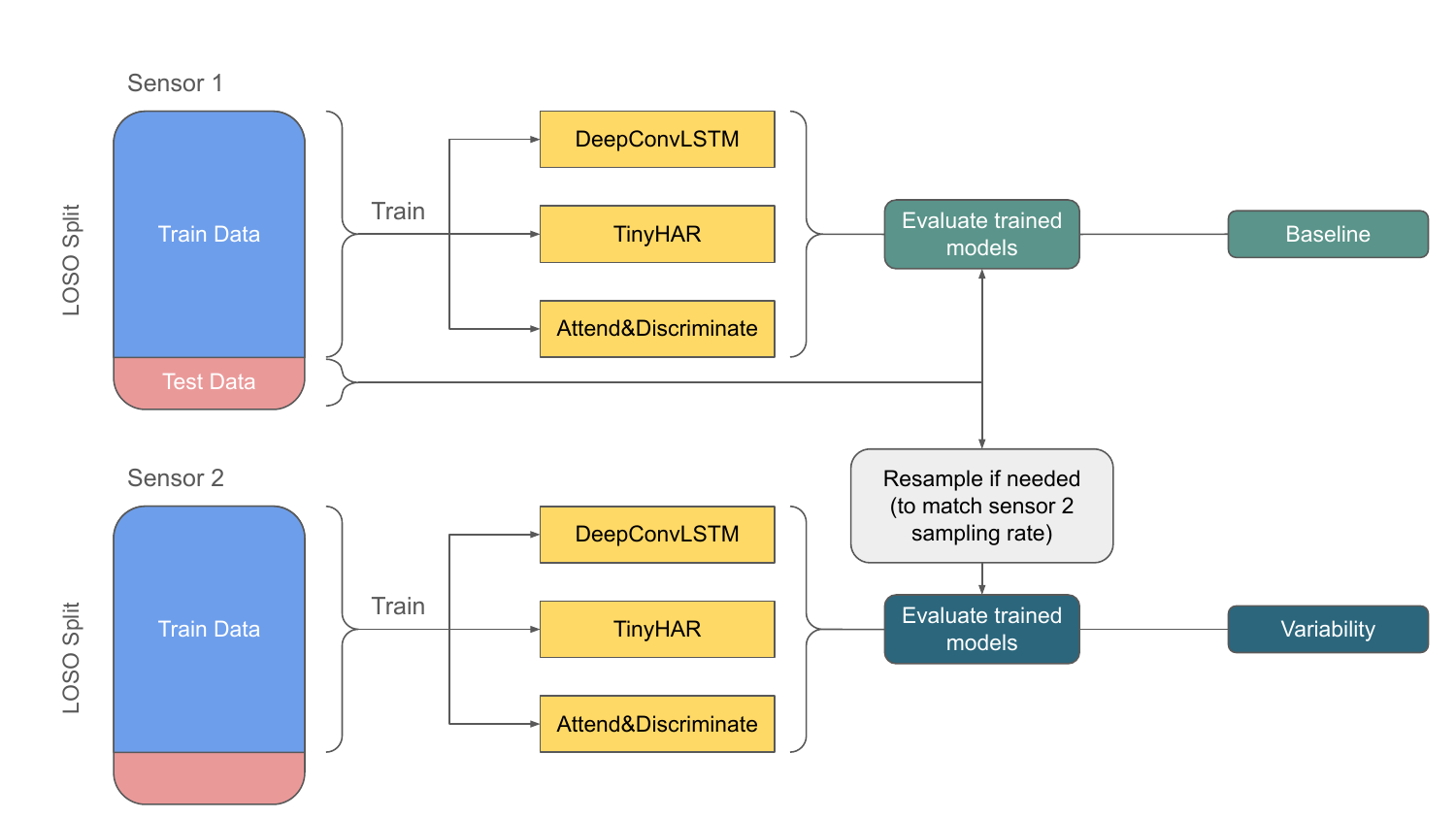}
    \caption{The process of evaluating the effect of variability using the HARVAR dataset. }
    \label{fig:harvar_process}
\end{figure}

\subsection{Model Training}
As mentioned, we evaluated three DL models: DeepConvLSTM, Attend\&Discriminate, and TinyHAR. 

We trained the models on a simple binary classification task: identifying whether or not the subject was walking. Walking was chosen because it is simple and, as shown by \citet{subvarmotion1}, complex motions may differ significantly between individuals. By focusing on this controlled walking activity, we aim to minimize the impact of motion variability, which is further mitigated through LOSO cross-validation during testing.


The models were trained for one sensor at a time, with the input being the 3-dimensional accelerometer data. The training data was normalized in isolation from the test data using standardization and then split into sliding windows of size 2 seconds. The sliding windows were shuffled and split into a 9:1 ratio of train and validation. No other form of pre-processing was used on the training or validation data to maintain originality. A weighted data loader was utilized during the training process as the samples of the "not-walking" class outweighed the "walking" class by a 2:1 ratio. Models were trained using a batch size of 256 over 150 epochs, with early stopping called after 15 epochs of no improvement over the validation set. The initial learning rate was set to 0.001, using learning rate annealing with patience of 7 epochs and a reduction factor of 0.1. The Adam optimizer was utilized, optimizing based on the CrossEntropy criterion.

Note that the model architecture remains consistent between all experiments. Still, since the Empatica and Bluesense sensors run at different sampling rates (as shown in Table~\ref{tab:sensors_list}), the model complexity varies depending on which sensor is used to train the model. This difference in model complexity is depicted in Table~\ref{tab:complexity_model}, and the model complexity is calculated as Multiple Accumulate Operations (MACs). Bluesense sensors have a sampling frequency of 100 Hz, and Empatica sensors have a sampling frequency of 64 Hz. Since we do not employ resampling before the training process, the input window size for models trained using Empatica sensors is 128, while the input window size for models trained using Bluesense sensors is 200. As detailed in the following subsection and shown in Figure~\ref{fig:harvar_process}, resampling will only be performed using interpolation during the testing phase when the train and test sensors have different sampling rates. Since every layer in the model depends on the input size, the models trained using Bluesense sensors are more complex than those trained with Empatica.

\begin{table}[]
\begin{tabular}{llll}
\toprule
Model                & Sensor           & Computational Complexity (MACs) & Parameters \\ 
\midrule
TinyHAR              & Bluesense        & $2.53 * 10^{6}$            & 24864      \\
DeepConvLSTM         & Bluesense        & $2.19 * 10^{7}$            & 227138     \\
Attend\&Discriminate & Bluesense        & $8.36 * 10^{7}$            & 297412     \\
TinyHAR              & EmpaticaEmbrace+ & $1.53 * 10^{6}$            & 24864      \\
DeepConvLSTM         & EmpaticaEmbrace+ & $1.33 * 10^{7}$            & 227138     \\
Attend\&Discriminate & EmpaticaEmbrace+ & $5.12 * 10^{7}$            & 297412     \\ 
\bottomrule
\end{tabular}
\caption{The computational complexity (in MACs) variance between the different models. The computational complexity depends on the model architecture and the sensor being used to train due to the difference in sensor sampling rates.}
\label{tab:complexity_model}
\end{table}

\subsection{Model Performance Evaluation}
\label{sec:model_performance}
Once the models are trained, they are tested using the data from the participants left out during the LOSO training process. Multiple tests use data from different sensors each time, achieving the train-test pairs depicted in Table~\ref{tab:experiments}. The F1 score is used to compare performance across various experimental settings. For device variability, due to the different sampling frequencies of the devices, the test data is interpolated to match the sampling frequency of the training data before being input into the model. For instance, in the variability setting of experiment 5 in Table~\ref{tab:experiments}, the input test data is upsampled, whereas in experiment 6, the test data is downsampled. This is because, in the experiment 5 variability scenario, the input data is upsampled from 64Hz to 100Hz, and in experiment 6, it is downsampled from 100Hz to 64Hz.

The significance of the difference in performance between the baseline and variability settings is tested using a T-test. The null hypothesis posits no difference between the baseline and variability scenarios. This hypothesis holds if the p-value of the T-test is greater than 0.05. Conversely, if there is a significant difference in performance, the p-value will be less than 0.05, indicated by *, less than 0.01 by **, and less than 0.001 by ***.

\subsection{Measuring variability with MMD}
\label{sec:mmd_var}
Maximum Mean Discrepancy (MMD) is a kernel-based statistical test used to determine the similarity between data distributions. We hypothesized that variability introduces a distribution shift in the data, contributing to the observed effects on the performance of DL HAR models. Our study employed the multiscale kernel for MMD with a bandwidth range of 0.2, 0.5, 0.9, 1.3, 1.5, and 1.6. A higher MMD value indicates a greater difference or shift in data distribution, whereas a smaller MMD value suggests a smaller shift. This approach helps quantify the impact of variability on the data distribution and, consequently, on model performance.

We calculate the MMD value between the train and test splits used to evaluate the DL HAR models, both with and without variability. We can determine whether this affects the model's performance by assessing the similarity or dissimilarity between the train and test datasets. This approach allows us to quantify how distribution shifts, introduced by variability, impact the effectiveness of the DL HAR models.

Time windows of 100 samples are randomly picked from the train and test set, and the MMD is calculated between them. We iterate this 50,000 times and calculate the average MMD. The MMD is only calculated over labeled data, not null or negative classes. So, in the HARVAR dataset, we only calculate the MMD over the walking class, not the activities labeled "not walking," as they are a heterogeneous mix of multiple activities.

\subsection{Measuring compounding effects of variability}

\begin{table}[]
\centering
\begin{tabular}{lllll}
\toprule
Exp. ID & Scenario                                              & Train Data & Test Data & Sensor \\ \midrule
1.      & Lab scenario, trained and tested with ideal data.     & Ideal      & Ideal     & RLA/LLA    \\
2.      & Trained with variable data and tested with ideal data & Self       & Ideal     & RLA/LLA    \\ \midrule
3.      & Trained and tested with variable data.                & Self       & Self      & RLA/LLA    \\
4.      & Lab trained and tested on variable data.             & Ideal      & Self      & RLA/LLA    \\ 

\bottomrule
\end{tabular}
\caption{Experiments to evaluate compound effects of variability using the REALDISP Dataset.}
\label{tab:compound-experiments}
\end{table}

While the HARVAR dataset allows us to measure the effects of each type of variability in isolation, real-life scenarios often involve a combination of these variabilities. We utilized the REALDISP~\cite{realdisp1, realdisp2} dataset to study this compounding effect. The REALDISP dataset highlights the combined effect of wearing variabilities by comparing data collected from wearable IMU sensors placed ideally by researchers (Ideal) and data from sensors worn unsupervised by participants (Self). It was collected from 16 participants over two iterations for each participant. In the first attempt (Self), the participants wore the sensors without the guidance of the researchers to mimic the real-life placement of consumers who wear smart devices with IMU sensors. The second time in the 'Ideal' scenario, the IMU sensors were attached to the participants by the researchers in an ideal position and orientation. 

In the 'Self' setting, the sensors' position and orientation differ from the 'Ideal' setting. The orientation can vary by as much as 180 degrees if worn upside down, as the sensors lack a reference for the "correct" orientation. Unlike the HARVAR dataset, the position variability here is more subtle, as it doesn't involve switching the sensor from one wrist to another. Instead, the variability comes from minor changes along the arm's length. For instance, depending on the participant's comfort, a sensor could be worn on the wrist or the forearm.

We conducted the experiments summarized in Table~\ref{tab:compound-experiments} to investigate how wearing variability, induced by the combined effects of position and orientation variability, impacts the performance of DL models. The selected scenarios represent various training and testing conditions commonly encountered in HAR model evaluation.

The first two scenarios compare the ideal case with a variability case. In the first scenario, both training and testing are conducted using 'Ideal' data, while in the second scenario, 'Self' data is used for training and 'Ideal' data for testing. These scenarios are analogous to experiments conducted using the HARVAR dataset, featuring a non-variability scenario (Ideal vs. Ideal) and a variability scenario (Self vs. Ideal). By comparing the performance drop between these two scenarios, we aim to understand the effect of compounded wearing variabilities, such as orientation and position. It's important to note that this evaluation differs from HARVAR in that HARVAR features controlled variability, where the variability is consistent across all participants. In contrast, the variability in the REALDISP dataset varies from participant to participant, as the 'Self' data depends on how each participant wears the sensor.

The next two scenarios simulate real-world conditions faced when training DL HAR models. The third scenario reflects training and testing data collected in unconstrained, real-world conditions, allowing variability in both training and testing. The fourth scenario represents a situation where a model is trained on lab-collected ideal data and then deployed in real-world settings where user-induced variabilities can affect performance (Ideal vs. Self).

Each of the four scenarios is run twice, using data from the right lower arm (RLA) and once from the left lower arm (LLA). This ensures that our results are not biased by any differences in data caused by the dominant and non-dominant arms.

MMD is calculated between the train and test sets, similar to the approach used with the HARVAR dataset. However, unlike HARVAR, where we only utilized two activities, the REALDISP dataset includes 33 different activities for classification. To calculate the MMD in this case, we compute the MMD between the train and test sets for each activity separately, then average these values to obtain a Mean MMD. This differs from the HARVAR approach, where MMD was calculated solely for the walking activity.

\section{Results and Discussion}
In this section, we present the results of our study. The section is organized into subsections, each exploring a research question. We begin by studying the impacts of data variability on model performance by isolating each type of variability in the HARVAR dataset. We then use the Maximum Mean Discrepancy Metric (MMD) to explain differences in performance across variabilities and participants. In Section~\ref{sec:real-disp-results}, we study the combined effects of variability using the REALDISP Dataset as a more realistic scenario. We then discuss the implications (Section~\ref{sec:implications}) and limitations (Section~\ref{sec:limitations}) of the study.

\subsection{Variability impacts on model performance}
\label{sec:results-variability}

\begin{figure}[ht]
     \subfloat[Orientation variability between BR2 and BR1 when the test sensor is BR1]{%
        \includegraphics[width=0.34\textwidth]{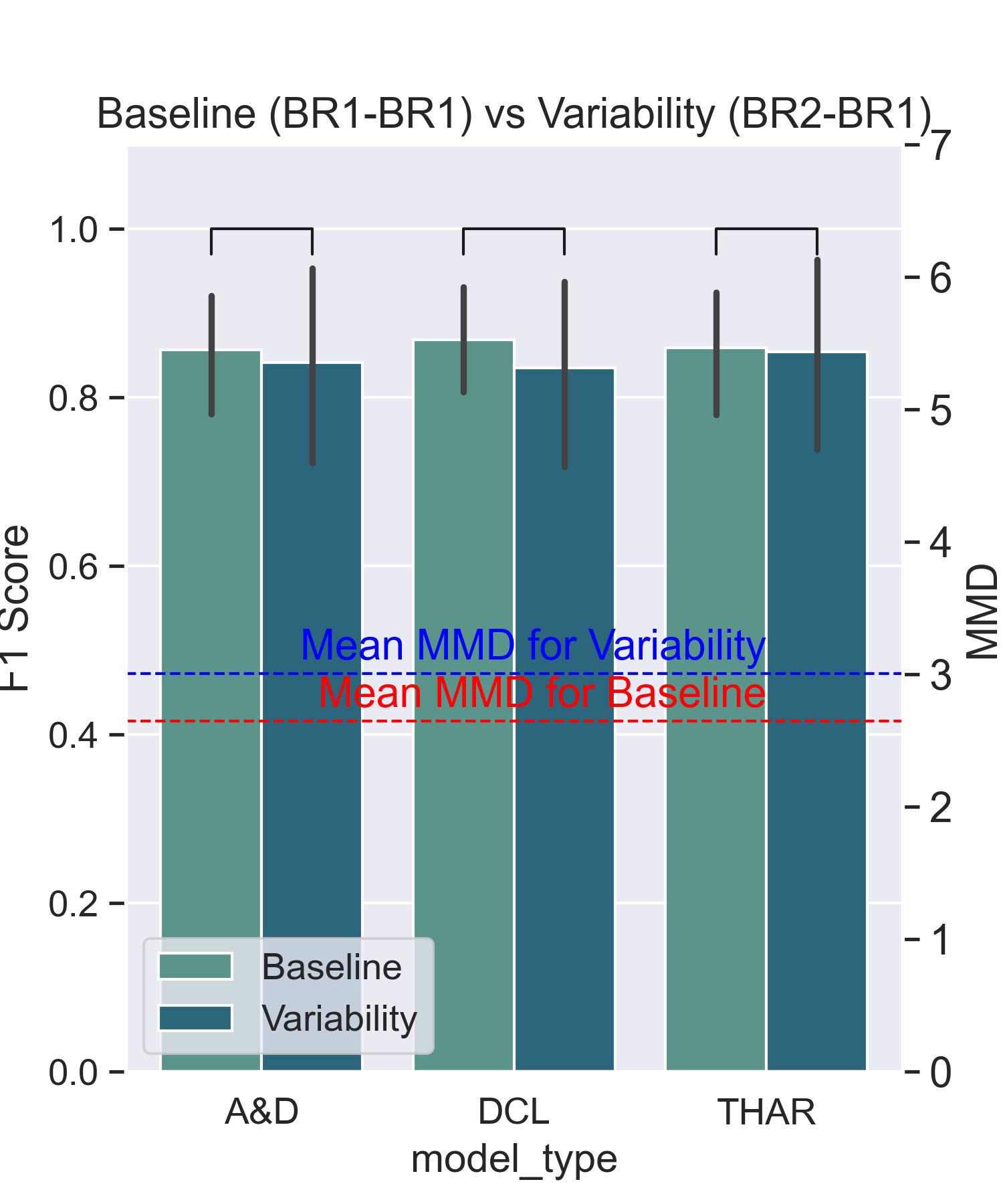}
        \label{fig:or2}
    }
    \centering
    \hspace{15pt}
    \subfloat[Orientation variability between BR2 and BR1 when the test sensor is BR2]{%
        \includegraphics[width=0.34\textwidth]{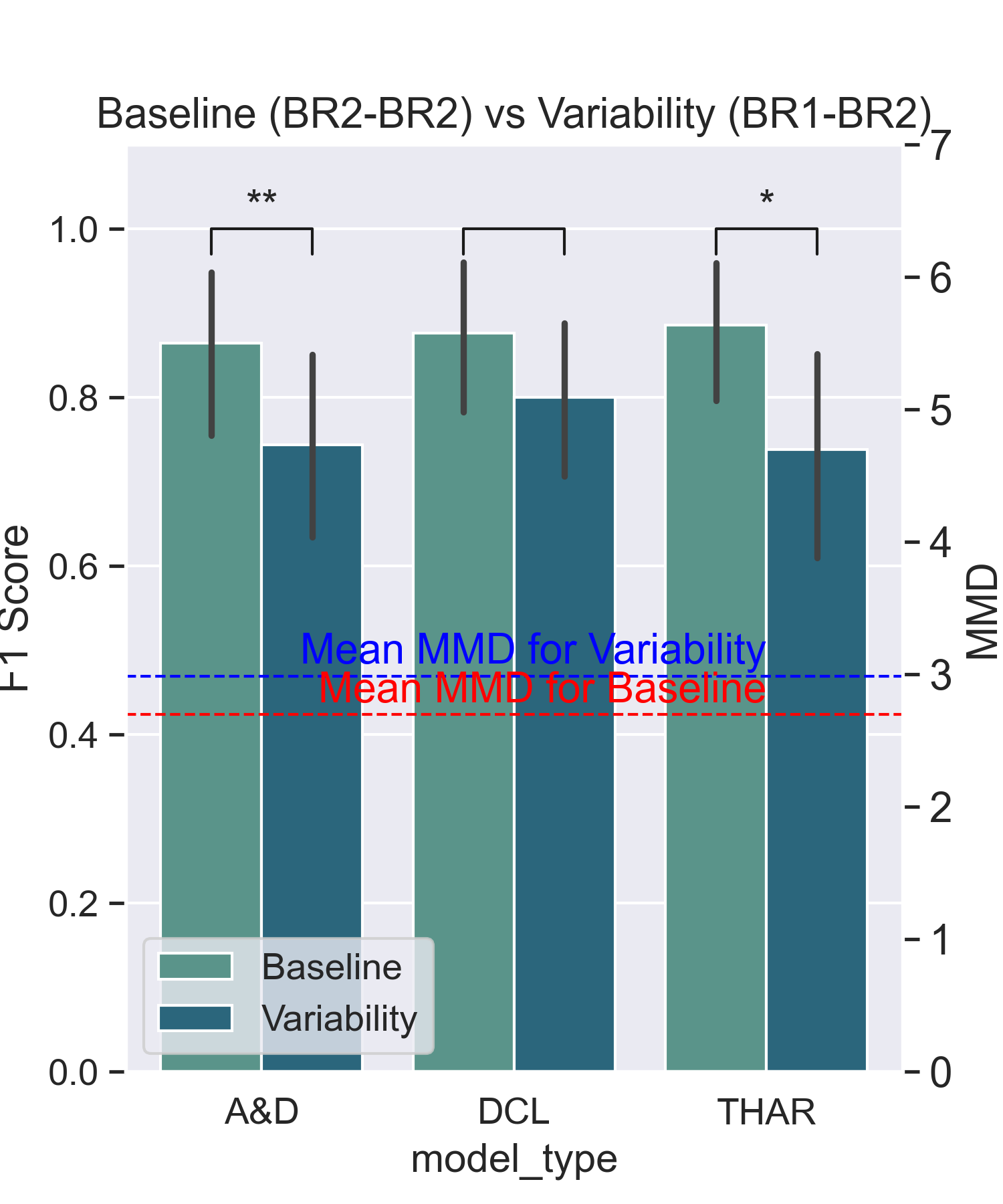}
        \label{fig:or1}
    }
    \caption{Performance changes due to Orientation Variability. We show the average F1 score and average MMD values for each DL HAR model in the two experiments. Light green bars represent the no variability setting of each experiment, and dark green bars represent the variability setting. Asterisks represent the p-value of a paired t-test (*: p-value <0.05, **: p-value <0.01, ***: p-value<0.001). Only two models in one experiment showed significant performance changes, but the F1-Score remains above 0.7.}
    \label{fig:or_parent}
\end{figure}

\begin{figure}[htbp]
    \centering
    \subfloat[EL - ER]{%
        \includegraphics[width=0.24\textwidth]{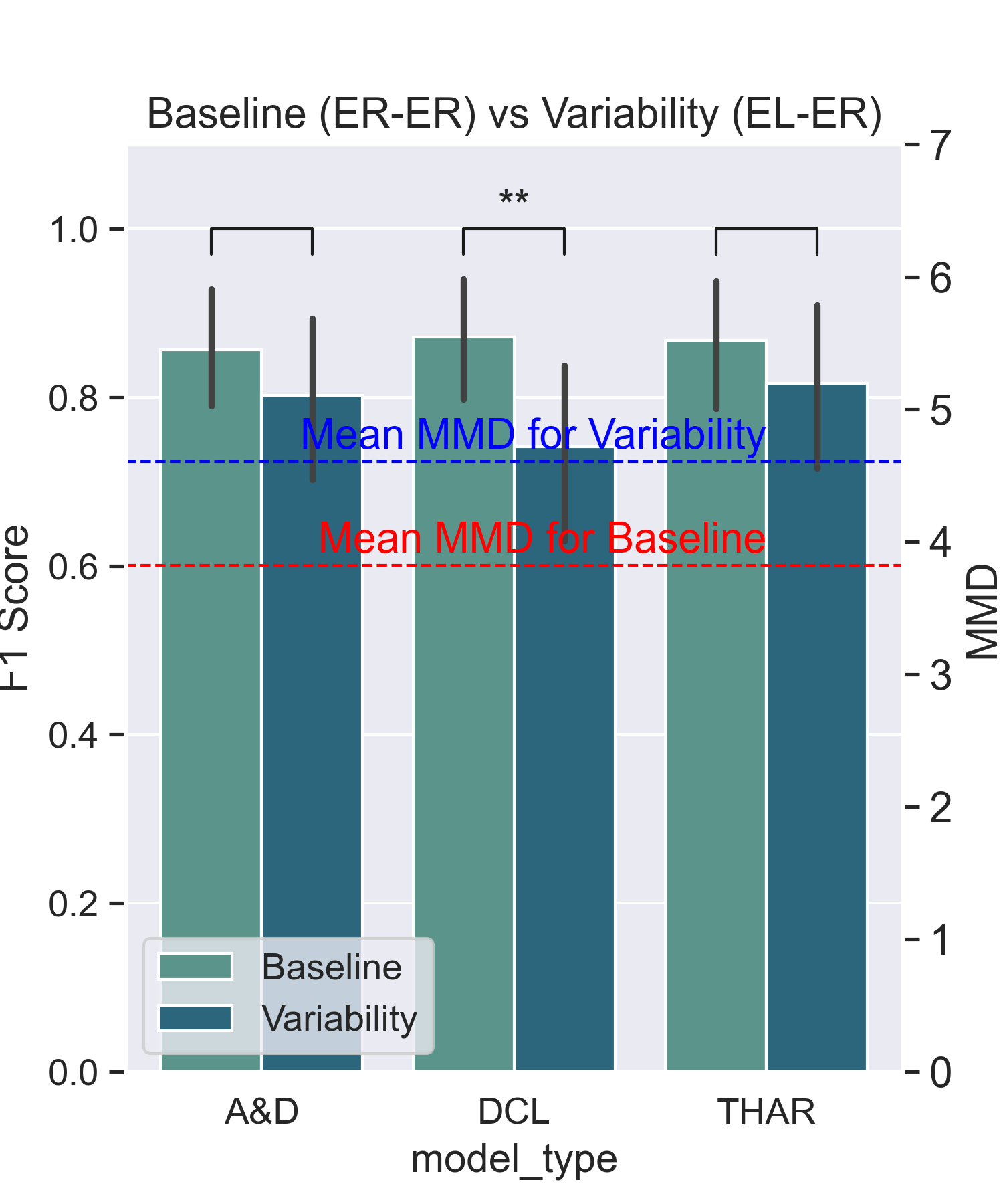 }
        \label{fig:po1}
    }
    \subfloat[ER - EL]{%
        \includegraphics[width=0.24\textwidth]{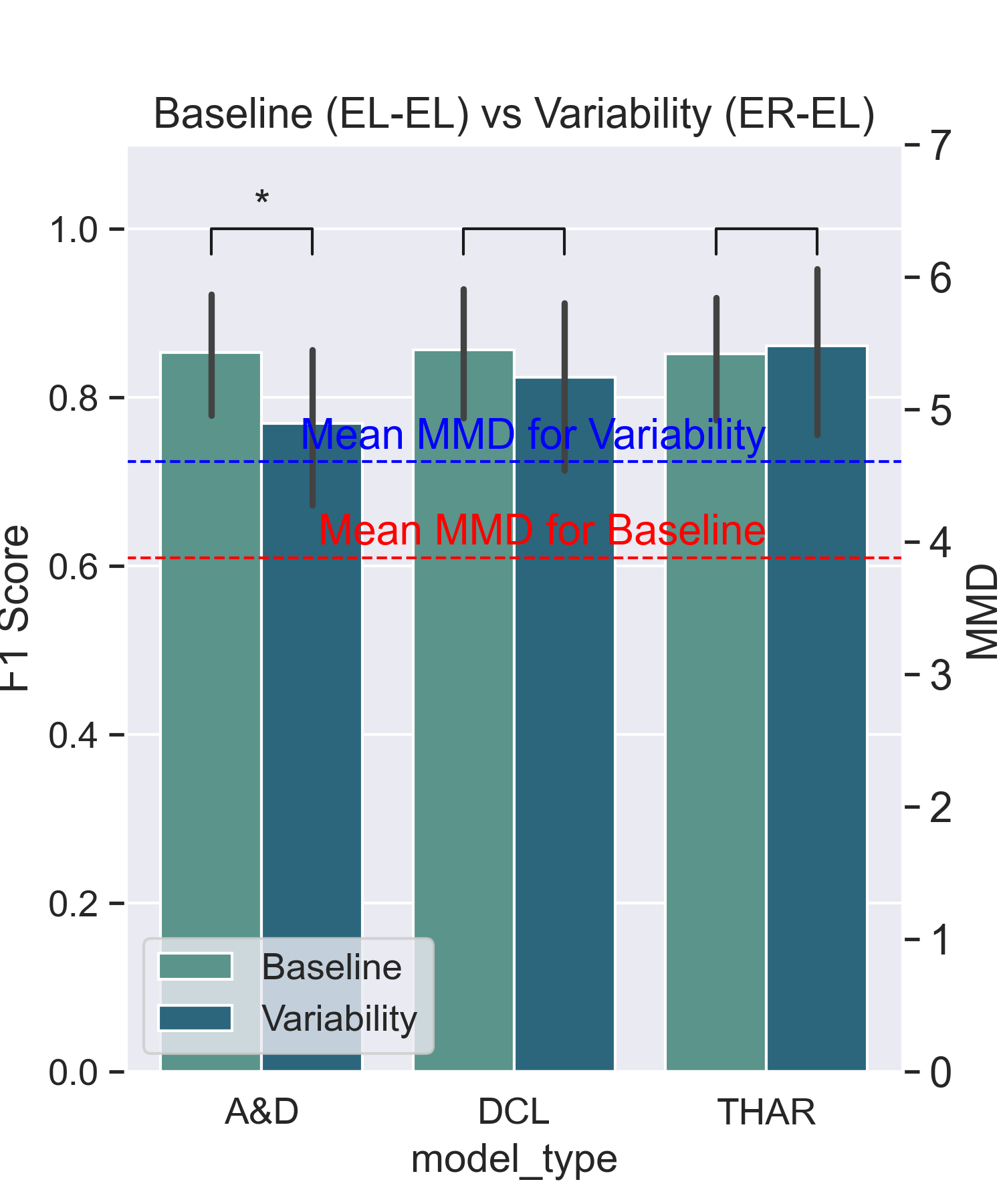 }
        \label{fig:po2}
    }
    \subfloat[BL - BR1]{%
        \includegraphics[width=0.24\textwidth]{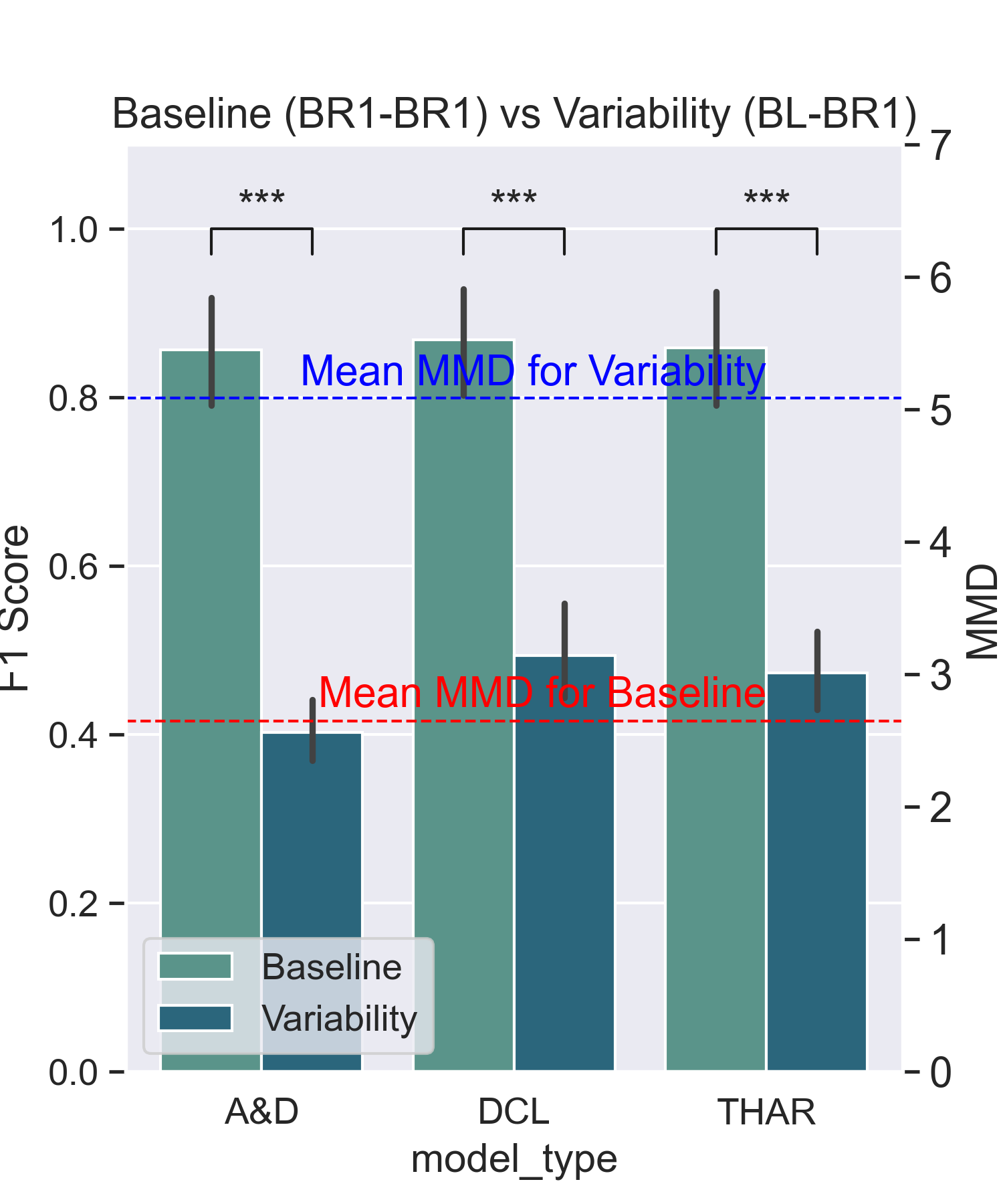}
        \label{fig:po3}
    }
    \subfloat[BR1 - BL]{%
        \includegraphics[width=0.24\textwidth]{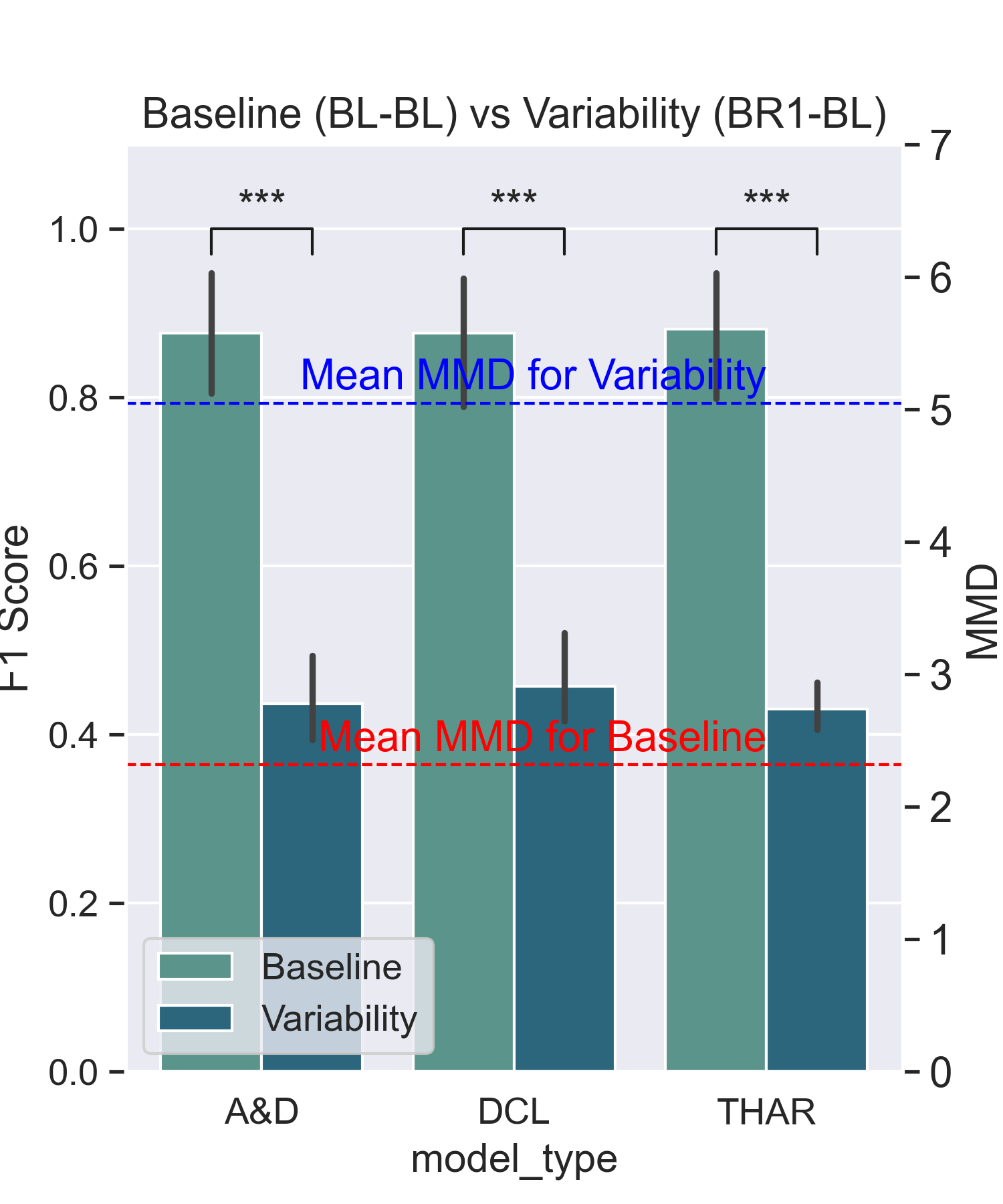}
        \label{fig:po4}
    }
    \caption{Performance changes due to Positional Variability. Bars represent the average F1 score for each DL HAR model and the lines represent the average MMD values of the settings. Light green bars represent the no variability setting; dark green bars represent the setting with variability. Asterisks represent the p-value of a paired t-test (*: p-value <0.05, **: p-value <0.01, ***: p-value<0.001). Significant performance changes were found for all models when BlueSense sensors were used but not for Empatica sensors.}
    \label{fig:po_parent}
\end{figure}

\begin{figure}[htbp]
        \subfloat[BL - EL]{%
        \includegraphics[width=0.24\textwidth]{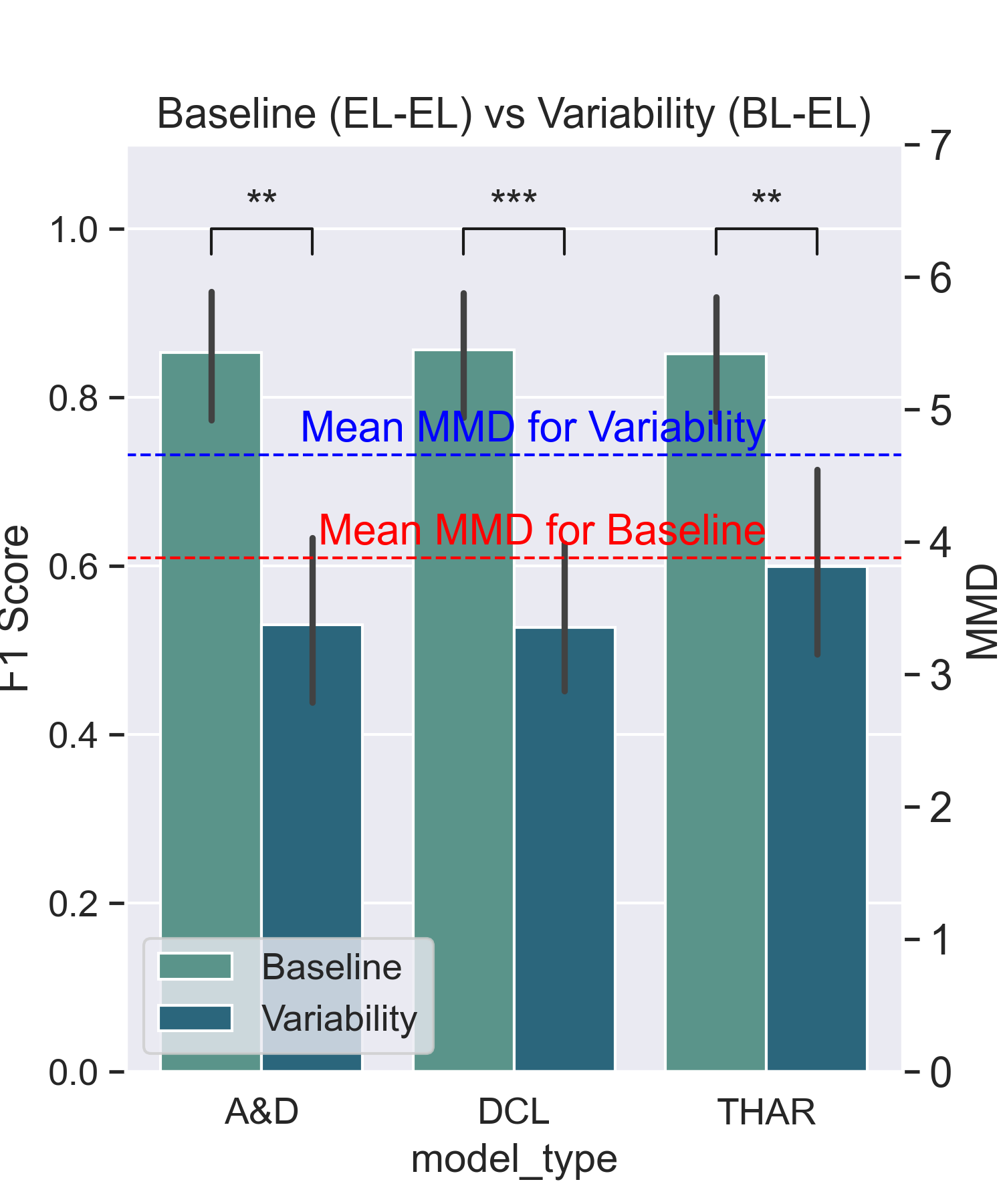 }
        \label{fig:dv1}
    }
    \subfloat[BR1 - ER]{%
        \includegraphics[width=0.24\textwidth]{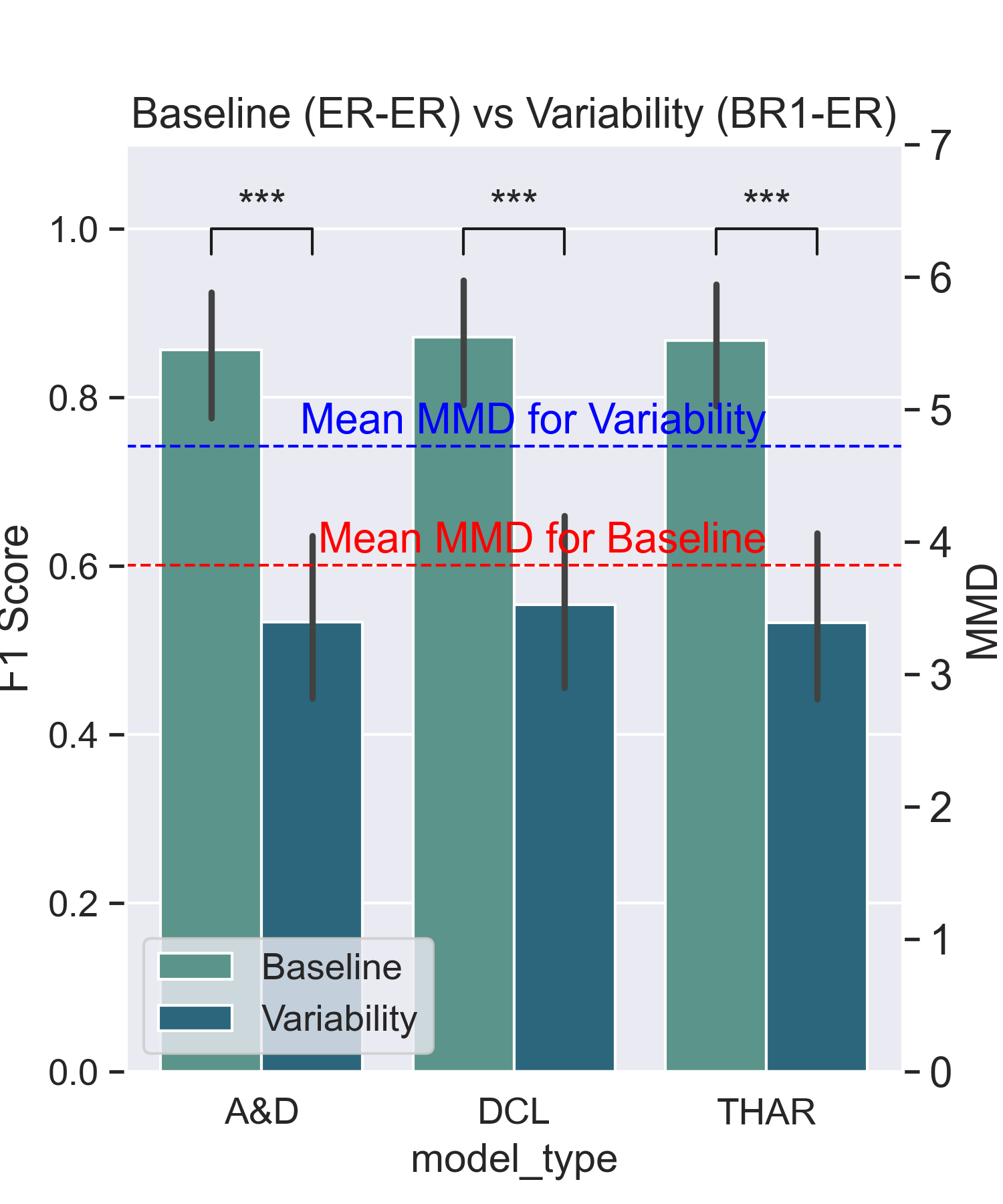 }
        \label{fig:dv2}
    }
    \centering
    \subfloat[EL- BL]{%
        \includegraphics[width=0.24\textwidth]{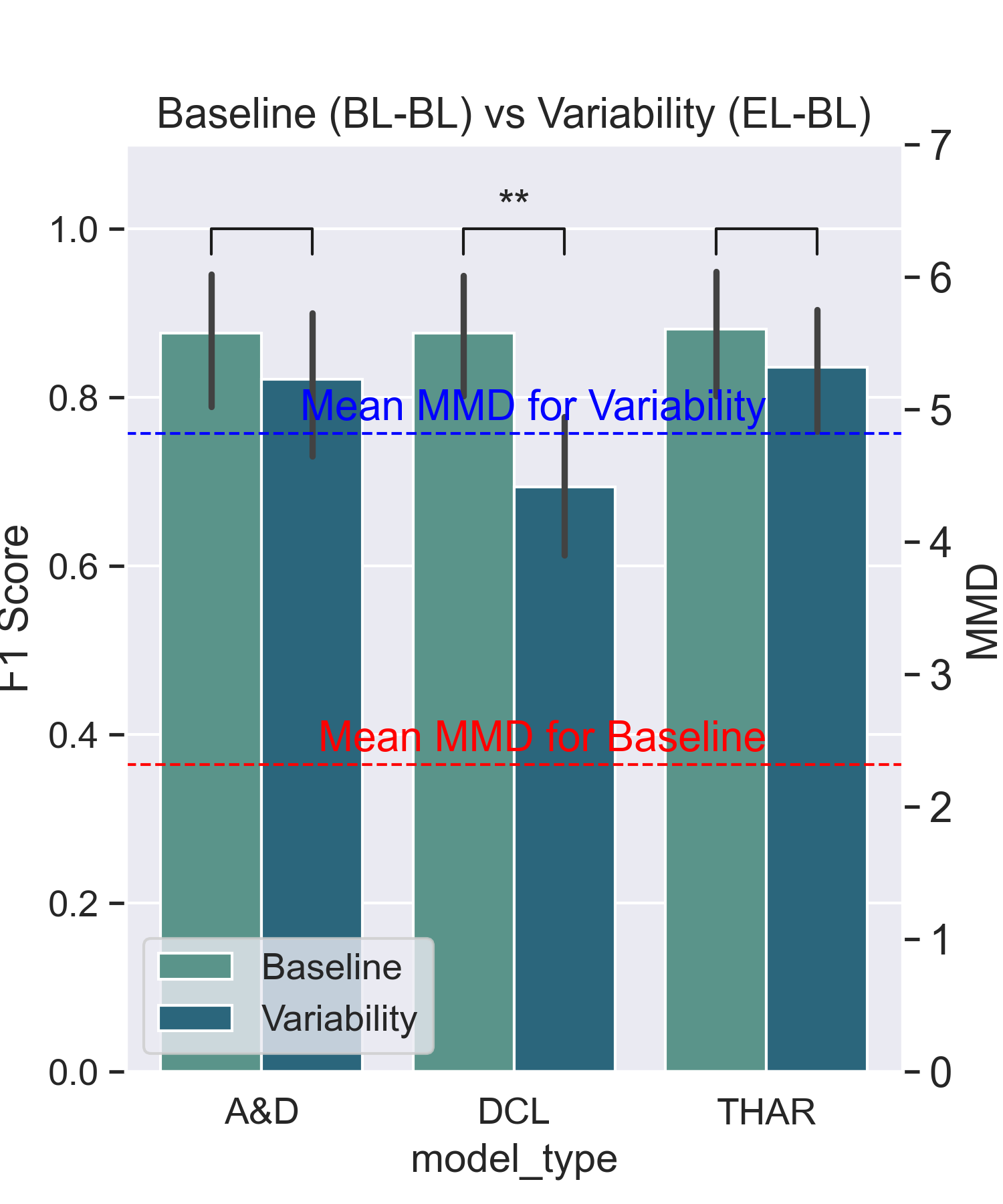 }
        \label{fig:dv3}
    }
    \subfloat[ER - BR1]{%
        \includegraphics[width=0.24\textwidth]{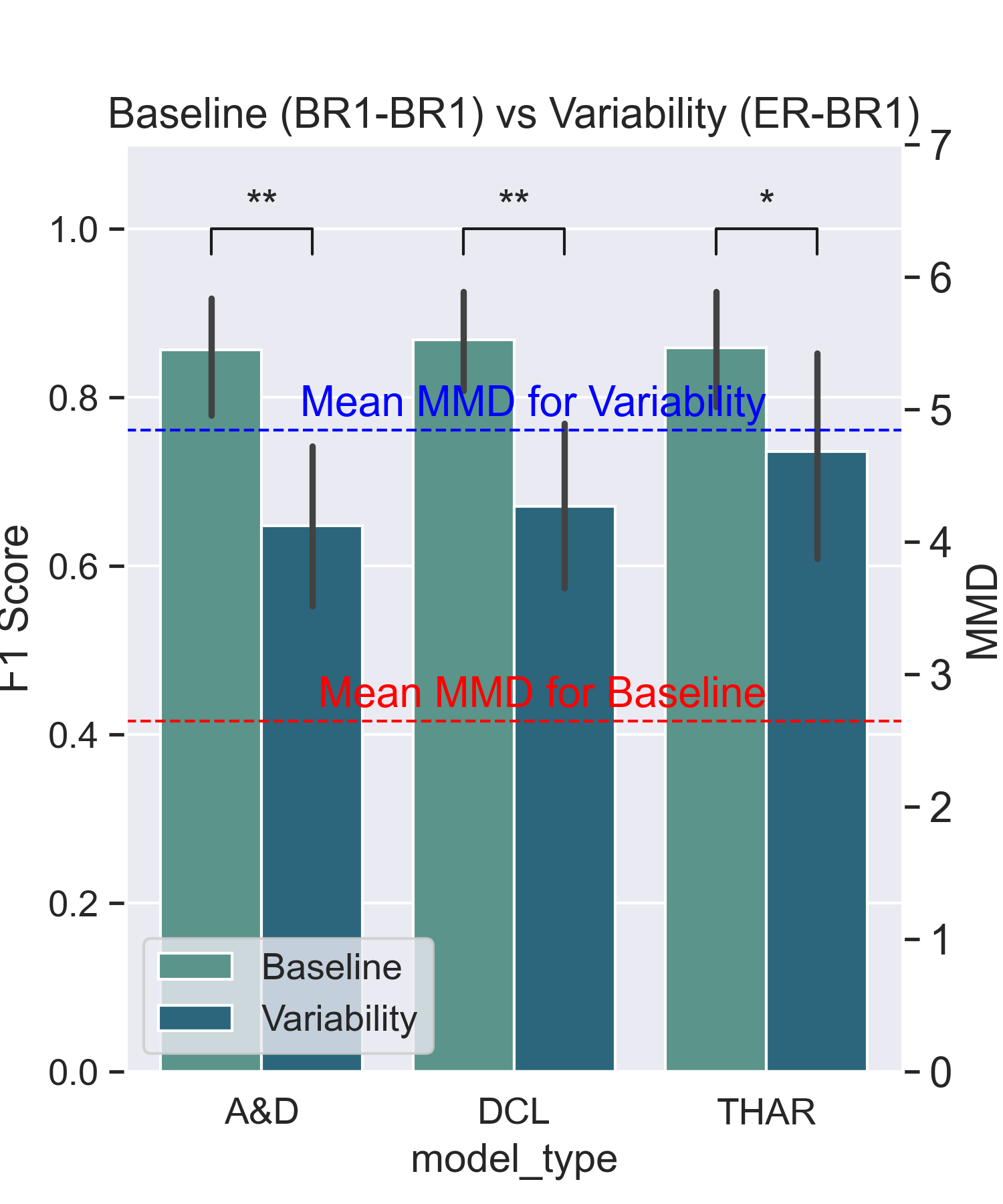}
        \label{fig:dv4}
    }
    \caption{Performance changes due to Device Variability. Bars represent the average F1 score for each DL HAR model, and the lines represent the setting's average MMD values. Light green bars represent the no variability setting; dark green bars represent the setting with variability. Asterisks represent the p-value of a paired t-test (*: p-value <0.05, **: p-value <0.01, ***: p-value<0.001). Significant differences in the performance were found in all but two cases. }
    \label{fig:dv_parent}
\end{figure}

We evaluated the impact of data variability on model performance by comparing the F1-Score difference on the baseline and variability settings. Figures~\ref{fig:or_parent}, \ref{fig:po_parent} and~\ref{fig:dv_parent} show the average and standard deviation of the F1-Score across all validation folds of each of the three evaluated models under no variability (dark green) and orientation, position, and device variability (light green) settings, respectively.
MMD values, also shown in these figures, will be explained in Section~\ref{sec:mmd-results}.

\textbf{Orientation variability} due to the rotation of an accelerometer along one of its axes was tested using the BR1 and BR2 sensors. These sensors have a 45-degree rotation difference but are both on the right wrist. Figure~\ref{fig:or_parent}, \ref{fig:po_parent}, and~\ref{fig:dv_parent} show the DL HAR model's mean F1 score for orientation, position, and device variability, respectively. The mean F1 score is calculated over the F1 score acquired from all participants during LOSO cross-validation.

Figure~\ref{fig:or_parent} depicts the model performance for experiments 7 and 8 in Table~\ref{tab:experiments}. In Figure~\ref{fig:or1}, we do not see any significant model performance changes due to orientation variability (p>0.05 in a paired t-test). In contrast, in Figure~\ref{fig:or2}, we see a significant drop in the performance of the Attend\&Discriminate model (p < 0.001) and in the performance of the TinyHAR model (p < 0.05). We do not see a significant drop in the performance of the DeepConvLSTM model (p > 0.05).

In both orientation variability experiments shown in Figure~\ref{fig:or_parent}, the performance of the baseline setting remains similar (F1 score 0.86) for all models regardless of the test sensor. However, in the variability scenario, we see a difference in performance between the two experiments:
\begin{enumerate}
    \item In Figure~\ref{fig:or1}, when the model is trained with the sensor BR2 (which is rotated 45 degrees) and tested with sensor BR1 (with no rotation), we see no significant drop in performance. In this variability experiment, the F1 score remains above 0.81 for all three DL models.
    \item In Figure~\ref{fig:or2} when the model is trained with the sensor BR1 (which has no rotation) and tested with sensor BR2 (with 45 degrees of rotation), we see a significant drop in performance for two DL HAR models (Attend\&Discriminate and TinyHAR). In this variability experiment, the F1 score is less than 0.8 for DeepConvLSTM and less than 0.75 for Attend\&Discriminate and TinyHAR.
\end{enumerate}





\textbf{Positional variability} was evaluated across four experiments, as shown in Figure~\ref{fig:po_parent}. The results varied depending on the sensors being used. Comparing Figures~\ref{fig:po1}~and~\ref{fig:po2} (Empatica), with Figures~\ref{fig:po3}~and~\ref{fig:po4} (Bluesense), we observe a greater performance drop due to positional variability when using Bluesense sensors (mean F1-Score difference of 0.45 and p < 0.001) than when using Empatica sensors (mean F1-Score difference of 0.12 and p-value close to 0.05). Since the type of variability is the same and the DL model architectures are unchanged, the larger drop in performance can be attributed to the differences between Bluesense and Empatica sensors and how positional variability causes a shift in their data distribution.

We note that the baseline performance of the DL models (indicated in light green) is consistent and independent of the sensor used, as shown throughout the experiments in Figure~\ref{fig:dv_parent}. 

The drop in performance due to position variability is inconsistent across models when empatica sensors are used. In Figure~\ref{fig:po1}, we only see DeepConVLSTM show a significant drop in performance (p < 0.001), whereas in Figure~\ref{fig:po2}, Attend\&Discriminate has the most significant drop in performance (p < 0.05). From the experiments done using the empatica sensors, we see that DL models, in general, can be robust against positional variability for simple activities such as walking.

\textbf{Device Variability}, shown in Figure~\ref{fig:dv_parent}, caused the most significant performance drop (p-value<0.001 for most cases) in the three DL HAR models compared to Position and Orientation Variability. The Device Variability experiments can be subdivided into two categories:
\begin{enumerate}
    \item Train bluesense and Test Empatica. Figures~\ref{fig:dv1} and~\ref{fig:dv2}
    \item Train Empatica and Test Bluesense. Figures~\ref{fig:dv3} and~\ref{fig:dv4}.
\end{enumerate}

We see that the performance drop due to Device Variability is larger for 'Train bluesense and Test Empatica' scenarios (mean F1-Score drop of 0.35) vs 'Train Empatica and Test Bluesense' scenarios (mean F1-Score drop of 0.17). 

\begin{figure}
    \centering
    \includegraphics[width=0.9\linewidth]{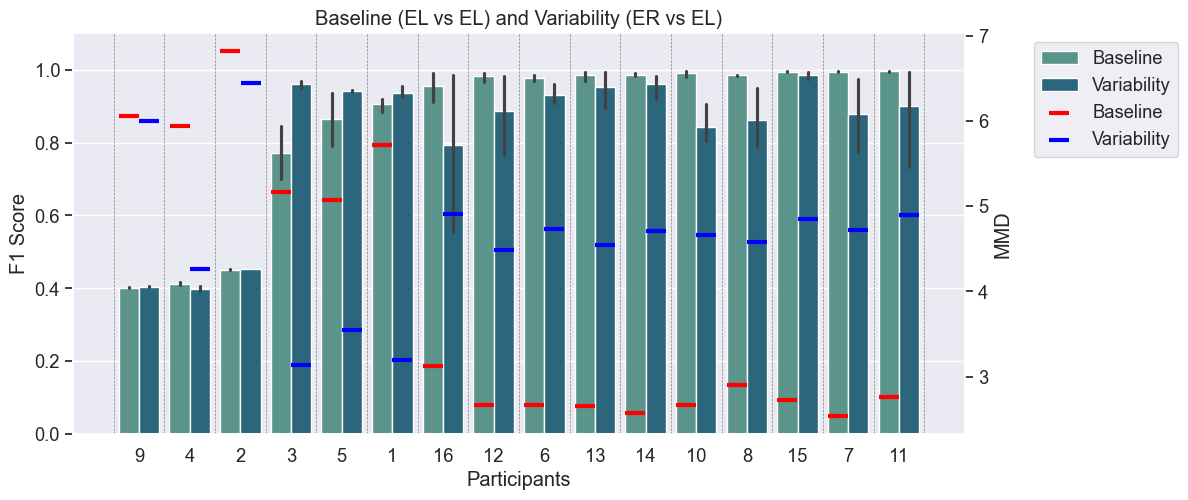}
    \caption{MMD of train vs test data and its relationship to the average F1-Score of the three evaluated models in the ER-EL experiment. This example depicts position variability where EL sensor data is used for testing. Dark green bars represent the F1-Score under variability, light green represents the baseline F1-Scores, and the blue and red points are their respective MMD values. The CV are arranged in ascending order of baseline F1 score.}
    \label{fig:mmd_example}
\end{figure}

With these results, we can observe how \noindent\textbf{variability reveals model performance nuances}. Throughout all the experiments conducted, we observed consistent mean performance across all models in the baseline experiments. These baseline experiments mimic the typical testing conditions for DL HAR models. Without variability, all DL-HAR models exhibited similar high performance. However, when we isolated a specific type of variability in our experiments, we observed varying performance among the different DL models, with some models exhibiting larger performance drops than others. These differences were inconsistent across experiments, with some models having larger differences in one experiment and smaller in another. Evaluating models with variability reveals their nuances and behavior under real-life conditions, demonstrating how they adapt to such changes. 
These results highlight the importance of testing DL HAR models under realistic conditions to better understand their robustness and adaptability.

\textbf{Subject variability} becomes evident when we examine the results at a granular level. Instead of just focusing on the mean performance, looking at each leave-one-subject-out cross-validation (CV) result reveals that the f1 scores for individual participants vary significantly. Figure~\ref{fig:mmd_example} illustrates an example of position variability by comparing sensors worn on the left and right wrists (Empatica-Left and Empatica-Right), detailing the f1 score per participant in ascending order of baseline f1 scores. Out of 16 participants, the first six deviate from the average trend. Participants 9, 4, and 2 exhibit very poor f1 scores of 0.4, indicating that the model's performance was comparable to making random guesses. Participants 3, 5, and 1 performed better in the variability scenario than in the baseline scenario. All other participants followed the expected trend, where baseline performance was higher than performance in the variability setting.



\subsection{Understanding Model Performance with MMD metric}
\label{sec:mmd-results}
As observed in the previous section, the effect of variability in model performance is unequal across types of variability, models, subjects and the selected test sensor. We hypothesized that the performance drop is related to the "amount of shift" in data distribution induced by the variability. To validate this, we use the Maximum Mean Discrepancy (MMD) metric, measuring the distance between two distributions. 

Figures~\ref{fig:or_parent},~\ref{fig:po_parent}, and~\ref{fig:dv_parent} depict the average MMD for the baseline setting with a red line, and the average MMD for the variability setting with a blue line. MMD is the same for all DL models in a given setting, as the train and test set are the same in each experiment. Since a higher MMD value indicates a greater shift in data distribution, higher MMD values represent more dissimilar distributions between the test and train data. 

\subsubsection{MMD to explain Orientation, Position and Device Variability}
We first study \textbf{differences in performance due to each type of variability}. Observing the average MMD values for all the experiments, it is apparent that the MMD is lower for the baseline than for the variability setting. This supports the hypothesis that variability causes a shift in data distribution. Moreover, the difference in MMD between the two settings is related to the difference in F1-Score, supporting the hypothesis that MMD is correlated with performance. 

In Figure~\ref{fig:or_parent}, the marginal difference between baseline and variability MMD values corresponds to the insignificant drop in performance due to orientation variability. Similarly, in Figures~\ref{fig:po1} and~\ref{fig:po2}, a small difference in mean MMD values aligns with a small drop in F1-Score.
In contrast, in Figures~\ref{fig:po3} and~\ref{fig:po4}, a greater difference in MMD values corresponds to a significant drop in DL model performance.
These observations indicate a relationship between MMD values and the performance change in DL HAR models due to variability. The MMD difference can explain the greater performance drop when Bluesense sensors are used as a test sensor compared to Empatica sensors in position variability scenarios. 

Figure~\ref{fig:dv_parent} presents a contrasting outcome to the positional variability observations in Figure~\ref{fig:po_parent}. Here, instead of a proportional drop in performance relative to the difference in mean MMD values, we observe that a smaller MMD difference is related to a larger performance drop in Figures~\ref{fig:dv1} and~\ref{fig:dv2}. Conversely, in Figures~\ref{fig:dv3} and~\ref{fig:dv4}, the MMD difference is larger, but the performance drop is less significant.

This discrepancy can be attributed to the differences in sampling rates between the sensors used in the experiments. Bluesense sensors sample at 100Hz, while Empatica sensors sample at 64Hz. This means that for the same 2-second time window, Bluesense sensors provide 200 samples, whereas Empatica sensors provide 128 samples. When a model is trained with Bluesense data (higher sampling rate) and tested with Empatica data (lower sampling rate), we must upsample the Empatica data. Upsampling does not introduce higher frequency features into the data, which might be essential for the model's accurate classification if trained with higher frequency information. On the other hand, if a model is trained with Empatica data and tested with Bluesense data, we downsample the Bluesense data. Downsampling removes high-frequency features from the test data, which the model, trained on lower-frequency data, does not rely on. Therefore, the performance drop is not as significant.

Another factor to consider is model complexity. Models trained with Bluesense sensors take longer inputs for the same time window than those trained with Empatica sensors, which means that the number of inputs in each layer is larger (Table~\ref{tab:complexity_model}). More Complex models may become highly specialized to the training data, which can increase their susceptibility to variability. This is because their complexity allows them to capture subtle details in the training data, which may not generalize well to data with different characteristics, leading to decreased performance when faced with variability. In contrast, less complex models might generalize better and thus perform more consistently under variability conditions. Further tests are required to confirm this and to explore whether more data can make the models more robust to variability. Nonetheless, it is important to remember that the amount of labeled sensor data available for HAR is usually small.

\subsubsection{MMD to explain Subject Variability}
We investigated the \textbf{differences in performance across participants} to highlight subject variability. We observed a high standard deviation in the F1-Score for each model, implying that each participant's performance depends on the participant's activity characteristics. To evaluate this, we measured the MMD for each cross-validation fold.  For example, in experiment 2, position variability, as shown in Figure~\ref{fig:mmd_example}.

For participants 9, 4, and 2, who achieved f1 scores around 0.4 (indicating the model struggled to distinguish between walking and not walking), their MMD values were notably higher. This aligns with the fact that these participants held onto support bars during the treadmill experiment, as shown in Table~\ref{tab:participantsInfo}, highlighting how slight variations in activity execution can heavily impact model performance.

Moreover, participants 3, 5, and 1 present an exception: their baseline MMD is higher than in the variability scenario. These participants performed better in the variability scenario but worse in the baseline scenario, which suggests that their test data in the variability setting was more similar to the training data compared to the baseline.

A consistent pattern emerges for participants 16, 12, 6, 13, 14, 10, 8, 15, 7, and 11: low MMD in the baseline setting and high MMD in the variability setting. This explains their higher performance in the baseline scenario and the drop in performance when variability was introduced.

\subsubsection{MMD correlation to F1 Score}
\begin{figure}
    \centering
    \includegraphics[width=0.45\linewidth]{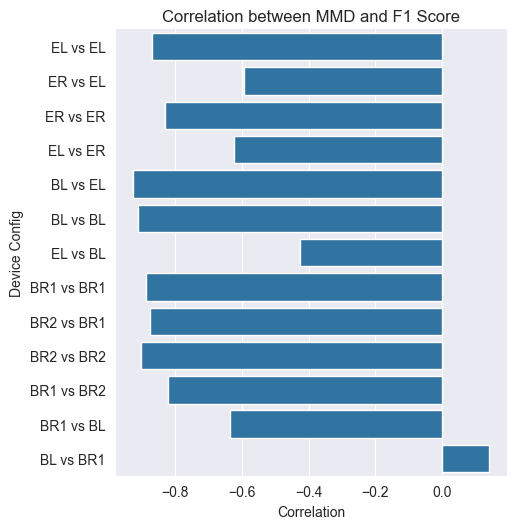}
    \caption{The correlation of the MMD values between train and test to the F1 score is mostly negative. Showing there is a negative correlation between the MMD and the Performance of a DL model.}
    \label{fig:corr}
\end{figure}

Upon calculating the correlation between the F1 score and the MMD between the train and test sets, we observed a negative correlation, as illustrated in Figure~\ref{fig:corr}. This supports our hypothesis that a relationship exists between the shift in data distribution and model performance. Almost all experiments demonstrated this negative correlation between the F1 score and MMD, further validating our hypothesis. 

However, an exception was found in the Bluesense-Left (BL) vs. Bluesense-Right (BR1) sensor experiment, where the correlation was closer to 0. This outlier can be attributed to the consistently poor performance of the models across all participants in the cross-validation, regardless of the MMD value. In scenarios where the model performs poorly overall, the impact of changes in MMD appears minimal.

We observed how introducing variability, whether from orientation, position, device, or subject, results in higher MMD in most cases. The MMD has shown how, for some participants, introducing variability helps the data become more similar to the train distribution, explaining why, in some cases, the performance increases when variability is introduced. This relationship underscores the impact of data distribution shifts on the performance of DL HAR models and highlights the importance of considering individual participant variability in model evaluation.

\subsection{Compounding Variability Effects in Real-Life Scenarios (REALDISP Case Study)}
\label{sec:real-disp-results}

\begin{figure}
    \centering
    \subfloat[Sensor used RLA]{
        \includegraphics[width=0.30\textwidth]{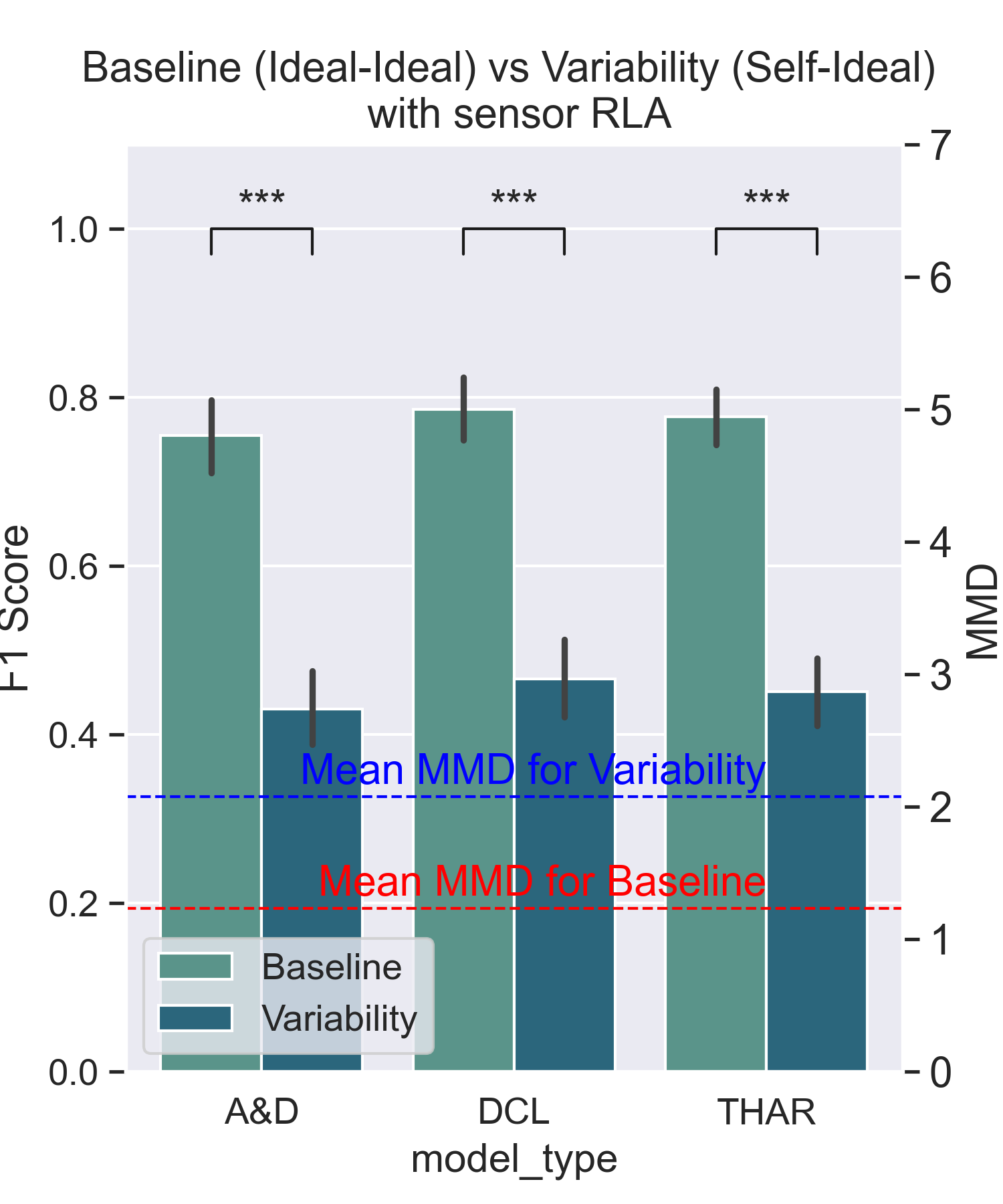}
        \label{fig:realdisp_RLA}
    }
    \centering
    \subfloat[Sensor used LLA]{
        \includegraphics[width=0.30\textwidth]{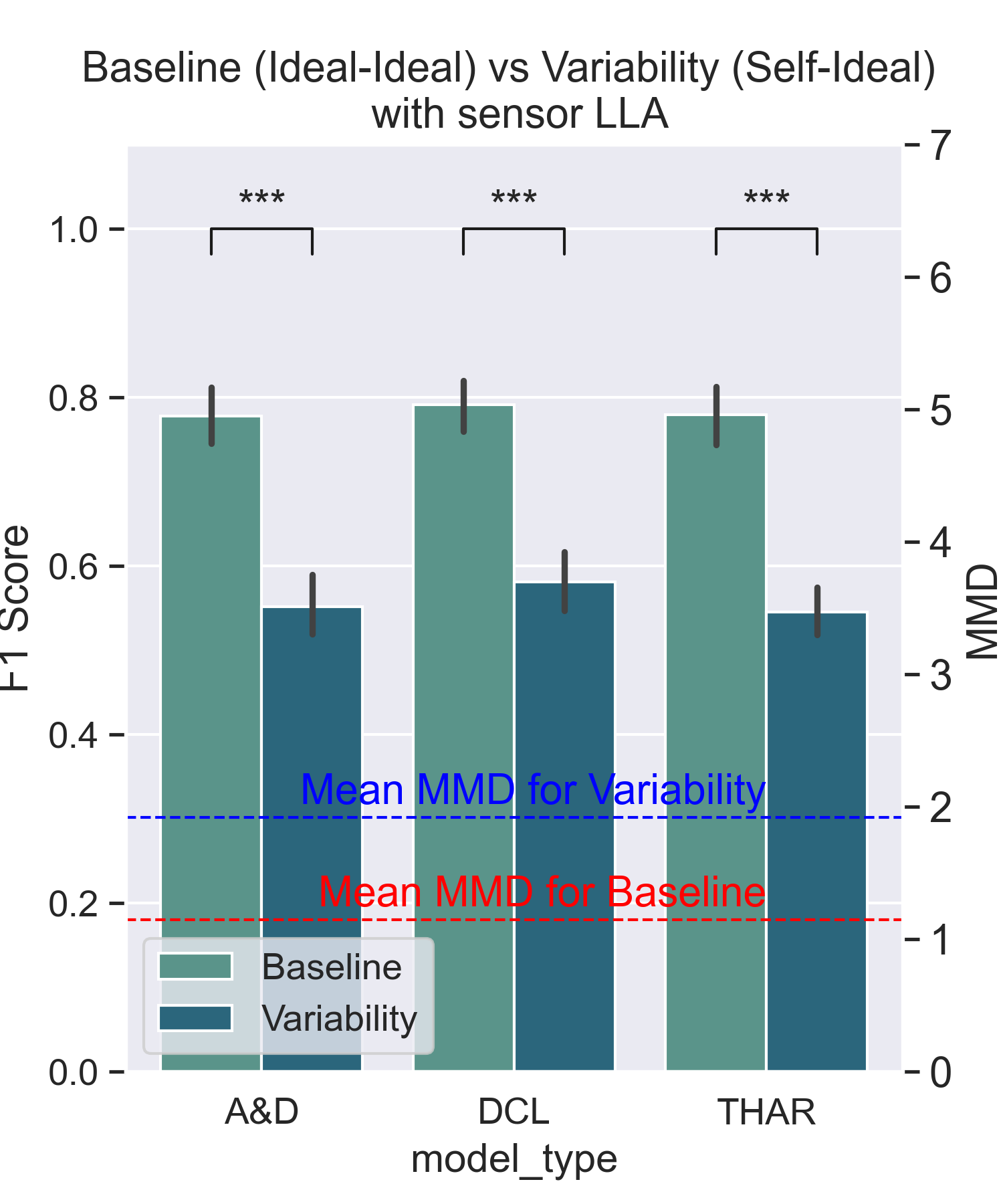}
        \label{fig:realdisp_LLA}
    }
    \caption{The mean F1 score for RLA and LLA sensors }
    \label{fig:realdisp}
\end{figure}

The results from the REALDISP dataset revealed a significant drop in performance for both RLA and LLA sensors due to the compounding effects of variability (p-value < 0.001), as shown in Figure~\ref{fig:realdisp}. Figure~\ref{fig:realdisp_RLA} illustrates the performance of DL models trained on data collected from the RLA sensor. Consistent with the findings from the HARVAR dataset, a higher MMD value corresponds to scenarios with poorer performance, while a lower MMD value corresponds to scenarios with better performance. Specifically, the MMD between the Ideal train and test data is much lower than the MMD between the Self-train and Ideal test data.

When analyzing the performance using the LLA sensor in figure~\ref{fig:realdisp_LLA}, we observe that in the Ideal vs. Ideal scenario, DL models perform similarly regardless of whether the RLA (F1 score 0.76) or LLA (F1 score 0.78) sensor data is used. However, in the Self vs. Ideal scenario, the LLA sensor outperforms the RLA sensor. The mean F1 score for the Self vs. Ideal scenario is 0.55 when using the LLA sensor, compared to 0.44 with the RLA sensor. This difference in performance is reflected in the MMD values: the MMD for LLA-Self vs. LLA-Ideal is 1.9, while RLA-Self vs. RLA-Ideal has an MMD of 2.05. These results further confirm that a lower MMD value corresponds to better model performance, while a higher MMD value indicates worse performance.

\begin{figure}
    \centering
    \includegraphics[width=0.7\linewidth]{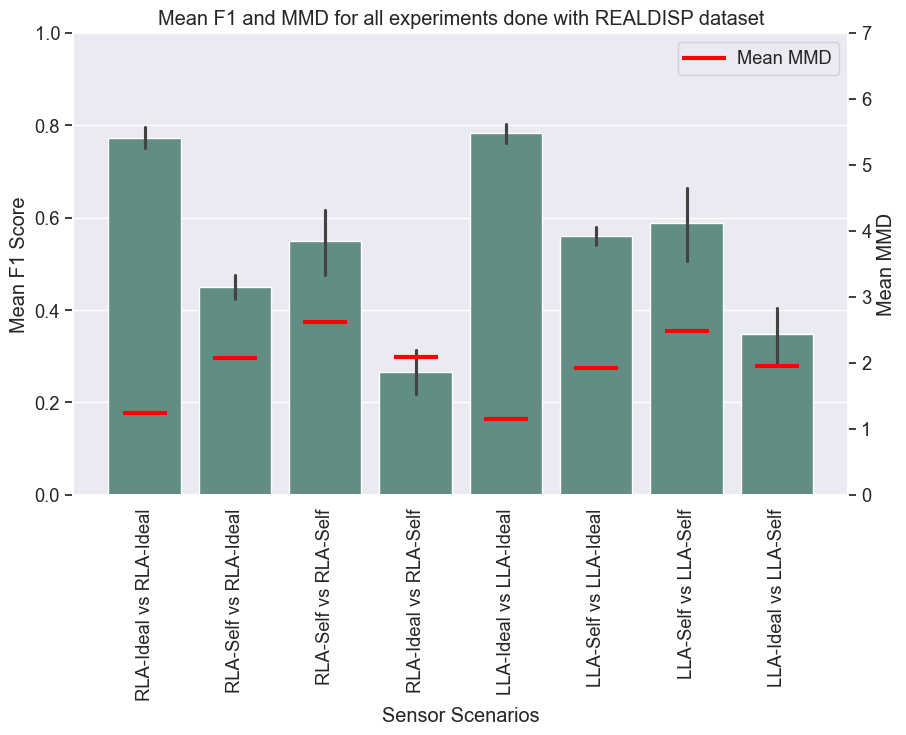}
    \caption{The mean F1 score and MMD values for all scenarios tested using the REALDISP dataset.}
    \label{fig:realdisp_all}
\end{figure}

Figure~\ref{fig:realdisp_all} shows the mean F1 score and MMD values for each scenario outlined in Table~\ref{tab:compound-experiments} for both RLA and LLA sensors. The best performance is observed in the scenario where both the training and testing data are collected under ideal conditions, which is expected since there is no variability to degrade the performance of the DL model. The poorest performance occurs when the model is trained on ideal data but tested on self-collected data. This indicates that any DL HAR model trained using lab-collected data would likely perform poorly when applied in real-world settings with wearing variabilities.

The MMD values reveal that the highest MMD occurs when training and testing data are of type self. While we've observed that higher MMD values generally correspond to lower performance, this trend does not hold in this case. The elevated MMD in the self vs. self scenario can be explained by the significant variability within the self data, as participants wear sensors in varied ways, often with sensors flipped across axes. This variability leads to a wider distribution shift, resulting in a higher MMD value. However, this diversity in the training data makes the model more robust to variability, leading to better generalization and performance in the self vs. self scenario.

In contrast, the ideal vs. self scenario suffers because the models trained on ideal data lack exposure to variability during training, making them vulnerable when tested under non-ideal conditions. Notably, despite having similar MMD values to the ideal vs. self scenario, the self vs. ideal scenario performs better. This can be attributed to the fact that when a DL HAR model is trained on diverse and variable data, it becomes more robust, resulting in improved performance even when tested on ideal data.

\subsection{Implications of Results}
\label{sec:implications}
This subsection highlights the major implications of the results of this study.

\subsubsection{Wearing variability and its implication on real-world scenarios}
Across the three types of variability studied in this paper, orientation variability caused the lowest performance drops across all models. This result suggests that models trained on IMU data from devices worn in fixed positions, such as smart glasses and earbuds, have more chances to generalize to multiple participants and environments, as the orientation variations that may occur will not significantly impact performance. 

Smartwatches are particularly vulnerable to the compounded effects of wearing variabilities, such as position and orientation. Experiments with the REALDISP dataset highlighted the significant impact of these variabilities on wrist-worn sensors, where orientation changes can be as extreme as a 180-degree flip across an axis. This drastic orientation shift further amplifies the effect on model performance when combined with position variability. Our findings also indicate that training models with a diverse dataset that includes a range of variabilities results in more robust performance, making them better suited for real-world applications.


\subsubsection{Device Variability has the most significant impact on performance}
Device variability had the most significant impact on the performance of DL models because it not only causes a shift in data distribution but also introduces differences in sampling frequency. These changes can affect the model size and necessitate resampling when using a device different from the one used in training. In addition, for this type of variability, MMD is insufficient to understand the variability.

Given the evolving wearable device industry, device variability is one of the main challenges to truly generalizable HAR models. Currently, different models for each device are required, which means updating models every time, which can be prohibitive if no data for the device has been collected. Researchers have utilized fine-tuning and domain adaptation methods in~\cite{crossHar} to overcome the effect of device variability in cross-dataset scenarios, where one dataset is used to train a model and another is used to test it.

Enhancing model robustness is crucial to address device variability, but it also requires careful data preprocessing and determining the optimal sampling rate for training the model. This would ensure that the model can generalize better across different devices.

\subsubsection{Subject Variability and the need for diverse training data}
The HARVAR results showed that variability in how individuals perform activities significantly impacts the performance of DL models. Human activity is inherently variable; these differences can change with age, demographics, and personal preferences. In our study, participants were asked to perform a simple treadmill walking task without specific instructions, leading some to hold the side bars while others did not. The reduced movement caused by holding the sidebars made it difficult for the models to classify walking accurately for those participants.

However, it's important to note that the training data for the DL models was not completely isolated from sidebar holding, as two participants in the training set also held onto the bars. Despite this, for CVs 9, 4, and 2, the sidebar holding data in train sets was outnumbered by non-side bar holding data in a ratio of 2:13. This highlights the models' bias towards the majority of the training data. To improve performance and generalizability for larger populations, datasets must either ensure better balance across activity variations or apply preprocessing techniques that give more weight to underrepresented data in the training set.

\subsubsection{Larger MMD correlate with smaller F1-Score, with limitations}
MMD serves as a useful metric to calculate the shift in data distribution. We observed a strong correlation between MMD and F1-Score, such as when MMD is large, F1-Score is low, and vice versa. 
Still, it sometimes fails to fully capture the impact of variability, as observed in the case of device variability. When changing devices alters the input shape to a model, MMD may not adequately explain the variability.

Additionally, MMD is a better metric when the F1 score is high, i.e., when the model's performance is good. However, beyond a certain threshold, when MMD is too high, changes in MMD stop reflecting in the changes to performance. As seen in Figure~\ref{fig:mmd_example}, spikes in MMD values for participants 2, 4, and 9 vary, but these three participants show an average F1 score of 0.41. On the other hand, when the performance is high, the difference in MMD shows a clear inverse relationship.

\subsubsection{No significant differences in performance change across the three models}
Statistical tests revealed no significant difference in performance between the three evaluated models. However, models with larger MACs tend to have bigger performance drops. High model complexity results from a larger input size due to a higher sampling rate or a larger network with more layers. 
We assume that increased complexity allows models to learn finer features, making them more prone to overfitting and less adaptable to changes. This aligns with previous results, such as the shallowLSTM~\cite{newdeepconvlst} network, which showed that using one less layer in the DCL model results in higher performance. 
In the face of non-significant performance changes, we recommend using lighter models, such as the TinyHAR model, which achieves similar performance and robustness with fewer parameters. 


\subsection{Study Limitations}
\label{sec:limitations}
This study isolated the effects of each type of variability in a binary classification task, while the DL models are capable of multiclass classification, as shown using the REALDISP dataset. Further studies are needed using multiclass classification with diverse activities in terms of motion and duration to better understand the robustness of the models. 

We evaluated the effects of variability in two datasets, each with 16/17 participants. While this number is small but similar to other public HAR datasets. However, the small size might not be enough to reveal significant differences across models and for some experiments. Increasing the number of participants can help reveal differences across the models, but larger datasets do not showcase the same type of variabilities observed in these two datasets.  

We studied wrist-worn sensor variabilities (orientation, position, and device). Future research should consider variability in other sensor placements, such as earbuds, chest-mounted sensors, and smart glasses. Device variability was only tested between two devices with 64Hz and 100Hz sampling frequencies, while many other devices with different noise levels and sensitivity ranges exist. As this type of variability showed the highest drops in performance, a deeper study on its effects and how to overcome it might be required. Other research~\cite {device_var_proof} have also found that the sampling rate of the train and test data should match for optimal performance.

Finally, we focused solely on accelerometer data, whereas many DL models are designed to combine multiple modalities, including gyroscope and magnetometer data, for HAR. We used only the accelerometer as it is the most common modality in many devices and has the lowest power consumption, making it preferable when possible. 

\section{Conclusion}
In this paper, we have studied three types of variability in three different DL-HAR models. We isolated each type of variability in our experiments, done with the HARVAR dataset specifically collected for this study. We evaluated the distribution and performance changes caused by position, orientation, and device changes. 


Our findings highlight that different types of variability affect DL HAR models in distinct ways. Orientation variability had the least impact on DL HAR models, whereas position and device variability resulted in significant performance drops. The impact of variability on DL HAR models also depends on the sensors used. For instance, position variability had a greater effect when using Bluesense sensors than Empatica sensors. 

Our study showed a direct relationship between changes in MMD values and the drop in DL model performance, emphasizing that a higher shift in data distribution (as indicated by MMD) corresponds to a lower performance (F1-Score). We recommend using MMD to predict potential performance drops when switching a model trained on a specific position, orientation, or device to another. Although MMD may not provide a complete picture, it is a useful metric for estimating performance changes.

We found that subject variability significantly impacted the performance of all three DL models. Variations in how activities are performed, especially for more complex activities than walking, can greatly affect model accuracy. This raises concerns about the reliability of DL models when tested on small, constrained datasets collected in controlled lab environments, where participants follow meticulously prescribed routines. Such settings may not capture the variability seen in real-world scenarios, questioning the generalizability of these models.




Using a sensor with a higher sampling rate increases the input size for DL HAR models, increasing their complexity. Models with higher MACs tend to perform worse in the presence of variability than models with lower MACs, as seen in Device Variability Scenarios. This suggests that a high sampling rate may lead the model to rely on nuanced high-frequency features, which diminishes the model's ability to handle variability effectively. Additionally, when comparing models like DeepConvLSTM, TinyHAR, and Attend\&Discriminate, we found that higher model complexity, as seen in DeepConvLSTM and Attend\&Discriminate, does not necessarily translate to better performance in handling variability. Therefore, we recommend using lighter models like TinyHAR, which consistently perform well despite variability.





This study aims to highlight the impact of various real-world variabilities on DL HAR models and examine their isolated effects. We analyzed the influence of three isolated variabilities on a simple activity like walking using the HARVAR dataset. Then, we demonstrated the effect of compounded variability across 33 different activities using the REALDISP dataset. A limitation of our work is that we did not isolate the effect of variability on activities other than walking, and our data was limited to 16 participants. Our findings suggest that incorporating a diverse range of variability in the training data enhances the robustness of DL models, as evidenced by the results from REALDISP. Utilizing MMD as a metric for data distribution shifts and the train-test pipeline developed in this research can enable future studies to evaluate the robustness of DL HAR models beyond ideally collected datasets.






\bibliographystyle{ACM-Reference-Format}
\bibliography{references}


\end{document}